\documentclass[10pt,conference,letterpaper]{IEEEtran}

\usepackage{caption}

\usepackage[dvipsnames]{xcolor}
\usepackage{algorithm}
\usepackage[noend]{algpseudocode}  

\makeatletter
\algrenewcommand\ALG@beginalgorithmic{\scriptsize}  
\algrenewcommand\algorithmiccomment[1]{\textcolor{OliveGreen}{// {\itshape #1}}}  
\newcommand{\CommentLine}[1]{
  \State \textcolor{OliveGreen}{// \textit{#1}}
}
\makeatother

\usepackage{textcomp}
\usepackage{graphicx}
\usepackage{epstopdf}
\usepackage{times}   
\usepackage{listings}
\usepackage{xspace}

\usepackage[colorlinks=true,citecolor=magenta,linkcolor=magenta,urlcolor=magenta]{hyperref}

\usepackage{booktabs}
\usepackage{multirow}

\usepackage[frozencache,cachedir=.]{minted}     

\usepackage{outlines}

\usepackage{soul}

\usepackage{amsmath}

\newcommand\subparagraph{%
  \@startsection{subparagraph}{5}
  {\parindent}
  {3.25ex \@plus 1ex \@minus .2ex}
  {-1em}
  {\normalfont\normalsize\bfseries}}
\makeatother
\usepackage{titlesec}
\let\subparagraph\relax 
\usepackage{titlesec}

\titlespacing*{\section}{0pt}{.5ex plus 1ex minus .2ex}{0ex plus .2ex}
\titlespacing*{\subsection}{0pt}{.5ex plus 1ex minus .2ex}{0ex plus .2ex}
\titlespacing*{\subsubsection}{0pt}{.5ex plus 1ex minus .2ex}{0ex plus .2ex}

\def\Snospace~{\S{}}

\newcommand{\ccc}[1]{}  

\newcommand{\joana}[1]{\ccc{\textcolor{orange}{JO: #1}}}

\newcommand{\KK}[1]{\ccc{\textcolor{purple}{KK: #1}}}
\newcommand{\julian}[1]{\ccc{\textcolor{cyan}{JS: #1}}}

\newcommand{\goner}[1]{} 

\newcommand{\sysname}{\textsc{Kaskade}\xspace}
\newcommand{\company}{Microsoft\xspace}
\newcommand{\ifequals}[3]{\ifthenelse{\equal{\detokenize{#1}}{\detokenize{#2}}}{#3}{}}
\newcommand{\case}[2]{}
\newenvironment{switch}[1]{\renewcommand{\case}{\ifequals{#1}}}{}
\newcommand{\vtype}[1]{\begin{switch}{#1}\case{spanner}{connector}\case{spanners}{connectors}\case{Spanner}{Connector}\case{Spanners}{Connectors}\case{sparsifier}{summarizer}\case{sparsifiers}{summarizers}\case{Sparsifier}{Summarizer}\case{Sparsifiers}{Summarizers}\end{switch}}

\newcommand{\speedupX}{50X\xspace}
\newcommand{\speedup}{50\xspace}

\newcommand{\mypara}[1]{\vspace{2mm}\noindent\textbf{#1}}

\renewcommand{\footnotesize}{\scriptsize}

\newmintedfile[prologcode]{prolog}{
  fontsize=\scriptsize,
}

\newmintedfile[sqlcode]{sql}{
  fontsize=\scriptsize,
}

\begin{document}
\bstctlcite{IEEEexample:BSTcontrol}


\title{Kaskade: Graph Views for Efficient Graph Analytics}

\author{\authorblockN{
  Joana M. F. da Trindade\authorrefmark{1},
  Konstantinos Karanasos\authorrefmark{2},
  Carlo Curino\authorrefmark{2},
  Samuel Madden\authorrefmark{1},
  Julian Shun\authorrefmark{1}}
 \authorblockA{
    \authorrefmark{1}MIT CSAIL \{jmf, jshun, madden\}@csail.mit.edu,
    \authorrefmark{2}Microsoft \{ccurino, kokarana\}@microsoft.com
  }
}

\maketitle


\begin{abstract}

Graphs are an increasingly popular way to model real-world entities and relationships between them, ranging from social networks to data lineage graphs and biological datasets. Queries over these large graphs often involve expensive sub-graph traversals and complex analytical computations. These real-world graphs are often substantially more structured than a generic vertex-and-edge model would suggest, but this insight has remained mostly unexplored by existing graph engines for graph query optimization purposes. Therefore, in this work, we focus on leveraging structural properties of graphs and queries to automatically derive materialized {\it graph views} that can dramatically speed up query evaluation.

We present \sysname, the first graph query optimization framework to exploit materialized graph views for query optimization purposes. \sysname employs a novel \emph{constraint-based view enumeration} technique that mines constraints from query workloads and graph schemas, and injects them during view enumeration to significantly reduce the search space of views to be considered. Moreover, it introduces a graph view size estimator to pick the most beneficial views to materialize given a query set and to select the best query evaluation plan given a set of materialized views. We evaluate its performance over real-world graphs, including the provenance graph that we maintain at \company to enable auditing, service analytics, and advanced system optimizations. Our results show that \sysname substantially reduces the effective graph size and yields significant performance speedups (up to \speedupX), in some cases making otherwise intractable queries possible.

\end{abstract}


\section{Introduction}
\label{s:intro}

Many real-world applications can be naturally modeled as graphs, including social networks~\cite{fbgraph,twitter}, workflow, and dependency graphs as the ones in job scheduling and task execution systems~\cite{halevy2016goods,dda-cidr-2017}, knowledge graphs~\cite{dbpedia,yago}, biological datasets, and road networks~\cite{netrepo}. An increasingly relevant type of workload over these graphs involves analytics computations that mix traversals and computation, e.g., finding subgraphs with specific connectivity properties or computing various metrics over sub-graphs. This has resulted in a large number of systems being designed to handle complex queries over such graphs~\cite{Yan2017,McCune2015}.

In these scenarios, graph analytics queries require response times on the order of a few seconds to minutes, because they are either exploratory queries run by users (e.g., recommendation or similarity search queries) or they power systems making online operational decisions (e.g., data valuation queries to control replication, or job similarity queries to drive caching decisions). However, many of these queries involve the enumeration of large subgraphs of the input graph, which can easily take minutes to hours to compute over large graphs on modern graph systems. To achieve our target response times over large graphs, new techniques are needed.

We observe that the graphs in many of these applications have an inherent structure: their vertices and edges have specific types, following well-defined schemas and connectivity properties. For instance, social network data might include users, pages, and events, which can be connected only in specific ways (e.g., a page cannot ``like'' a user), or workload management systems might involve files and jobs, with all files being created or consumed by some job. As we discuss in \autoref{sec:example}, the provenance graph that we maintain at \company has similar structural constraints. However, most existing graph query engines do not take advantage of this structure to improve query evaluation time.

At the same time, we notice that \emph{similar queries are often run repeatedly} over the same graph. Such queries can be identified and materialized as views to avoid significant computation cost during their evaluation. The aforementioned \emph{structural regularity} of these graphs can be exploited to efficiently and automatically derive these materialized views. Like their relational counterparts, such {\it graph views} allow us to answer queries by operating on much smaller amounts of data, hiding/amortizing computational costs and ultimately delivering substantial query performance improvements of up to \speedupX in our experiments on real-world graphs. As we show, the benefits of using graph views are more pronounced in \emph{heterogeneous} graphs\footnote{By heterogeneous, we refer to graphs that have more than one vertex types, as opposed to homogeneous with a single vertex type.} that include a large number of vertex and edge types with connectivity constraints between them.

\subsection{Motivating example}
\label{sec:example}

\begin{figure}[t!]
\centering\includegraphics[width=0.4\textwidth]{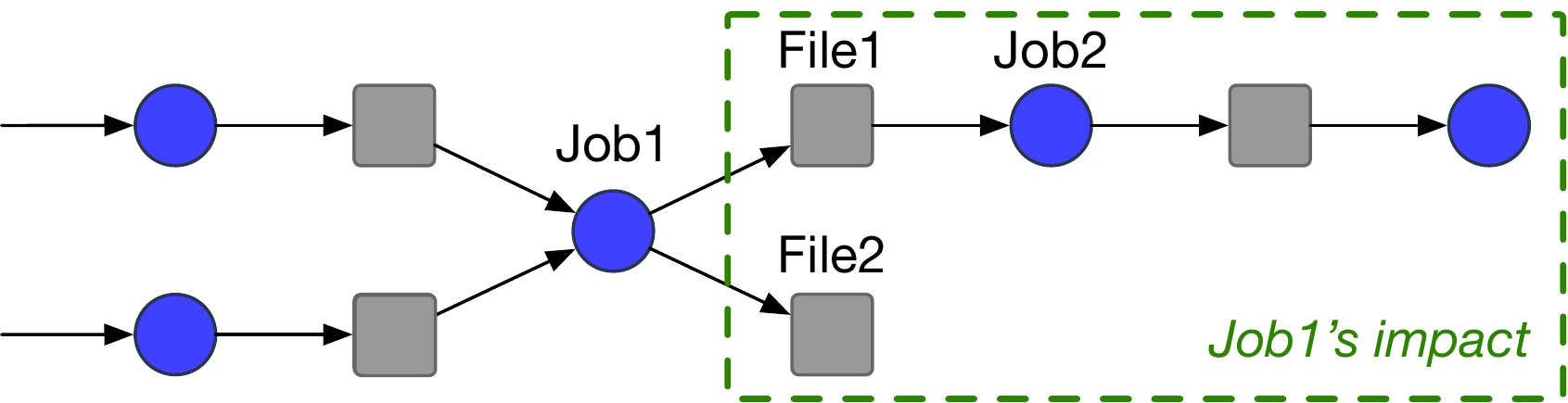}
\caption{Running example of query over a heterogeneous network: the ``blast radius'' impact for a given job in a data lineage graph (blue circles correspond to jobs; gray squares to files).}
\label{fig:blast_radius}
\end{figure}

At \company, we operate one of the largest data lakes worldwide, storing several exabytes of data and processing them with hundreds of thousands of jobs, spawning billions of tasks daily~\cite{hydra}. Operating such a massive infrastructure requires us to handle data governance and legal compliance (e.g., GDPR~\cite{gdpr}), optimize our systems based on our query workloads, and support metadata management and enterprise search for the entire company, similar to the scenarios in~\cite{halevy2016goods,dda-cidr-2017}. A natural way to represent this data and track datasets and computations at various levels of granularity is to build a {\em provenance graph} that captures data dependencies among jobs, tasks, files, file blocks, and users in the lake. As discussed above, only specific relationships among vertices are allowed, e.g., a user can submit a job, and a job can read or write files.

To enable these applications over the provenance graph, we need support for a wide range of {\it structural} queries. Finding files that contain data from a particular user or created by a particular job is an anchored graph traversal that computes the reachability graph from a set of source vertices, whereas detecting overlapping query sub-plans across jobs to avoid unnecessary computations can be achieved by searching for jobs with the same set of input data. Other queries include label propagation (i.e., marking privileged derivative data products), data valuation (i.e., quantifying the value of a dataset in terms of its ``centrality'' to jobs or users accessing them), copy detection (i.e., finding files that are stored multiple times by following copy jobs that have the same input dataset), and data recommendation (i.e., finding files accessed by other users who have accessed the same set of files that a user has).

We highlight the optimization opportunities in these types of queries through a running example: the \emph{job blast radius}. Consider the following query operating on the provenance graph: \emph{``For every job $j$, quantify the cost of failing it, in terms of the sum of CPU-hours of (affected) downstream consumers, i.e., jobs that directly or indirectly depend on $j$'s execution.''} This query, visualized in \autoref{fig:blast_radius}, traverses the graph by following read/write relationships among jobs and files, and computes an aggregate along the traversals. Answering this query is necessary for cluster operators and analysts to quantify the impact of job failures---this may affect scheduling and operational decisions.

By analyzing the query patterns and the graph structure, we can optimize the job blast radius query in the following ways. First, observe that the graph has structural connectivity constraints: jobs produce and consume files, but there are no file-file or job-job edges. Second,  not all vertices and edges in the graph are relevant to the query, e.g., it does not use vertices representing tasks. Hence, we can prune large amounts of data by storing as a view only vertices and edges of types that are required by the query. Third, while the query traverses job-file-job dependencies, it only uses metadata from jobs. Thus, a view storing only jobs and their ($2$-hop) relationships to other jobs further reduces the data we need to operate on and the number of path traversals to perform. A key contribution of our work is that these {\it views} of the graph can be used to answer queries, and if materialized, query answers can be computed much more quickly than if computed over the entire graph.

\subsection{Contributions}

Motivated by the above scenarios, we have built \sysname, a graph query optimization framework that employs graph views and materialization techniques to efficiently evaluate queries over graphs. Our contributions are as follows:

\mypara{Query optimization using graph views.} We identify a class of graph views that can capture the graph use cases we discussed above. We then provide algorithms to perform view selection (i.e., choose which views to materialize given a query set) and view-based query rewriting (i.e., evaluate a query given a set of already materialized views). To the best of our knowledge, this is the first work that employs graph views in graph query optimization. 

\mypara{Constraint-based view enumeration.} Efficiently enumerating candidate views is crucial for the performance of view-based query optimization algorithms. To this end, we introduce a novel technique that mines constraints from the graph schema and queries. 
It then leverages view templates expressed as inference rules to generate candidate views, injecting the mined constraints at runtime to reduce the search space of views to consider. The number of views our technique enumerates is further lowered when using the query constraints. 

\mypara{Cost model for graph views.} Given the importance of path traversals in graph queries, we introduce techniques to estimate the size of views involving such operations and to compute their creation cost. Our cost model is crucial for determining which views are the most beneficial to materialize. Our experiments reveal that by leveraging graph schema constraints and associated degree distributions, we can estimate the size of various path views in a number of real-world graphs reasonably well.

\mypara{Results over real-world graphs.} We have incorporated all of the above techniques in our system, \sysname, and have evaluated its efficiency using a variety of graph queries over both heterogeneous and homogeneous graphs. \sysname is capable of choosing views that, when materialized, speed up query evaluation by up to \speedupX on heterogeneous graphs.

\vspace{1mm}

The remainder of this paper is organized as follows. \autoref{s:overview} provides an overview of \sysname. \autoref{s:prelims} describes our graph data model and query language, and introduces the notion of {\em graph views} in \sysname. \autoref{s:venum} presents \sysname's constraint-based approach of enumerating graph views, whereas \autoref{s:viewops} discusses our graph view cost model and our algorithms for view selection and view-based query rewriting. \autoref{s:graph-views-detailed} gives more details on graph views (definitions and examples). \autoref{s:experiments} presents our experimental evaluation. Finally, \autoref{s:related} gives an overview of related work, and  \autoref{s:conclusions} provides our concluding remarks.

\section{Overview}
\label{s:overview}

\sysname is a graph query optimization framework that materializes graph views to enable efficient query evaluation.  As noted in~\autoref{s:intro}, it is designed to support complex enumeration queries over large subgraphs, often involving reporting-oriented applications that repeatedly compute filters and aggregates, and apply various analytics over graphs.

\begin{figure}
\begin{center}
\includegraphics[width=0.95\columnwidth]{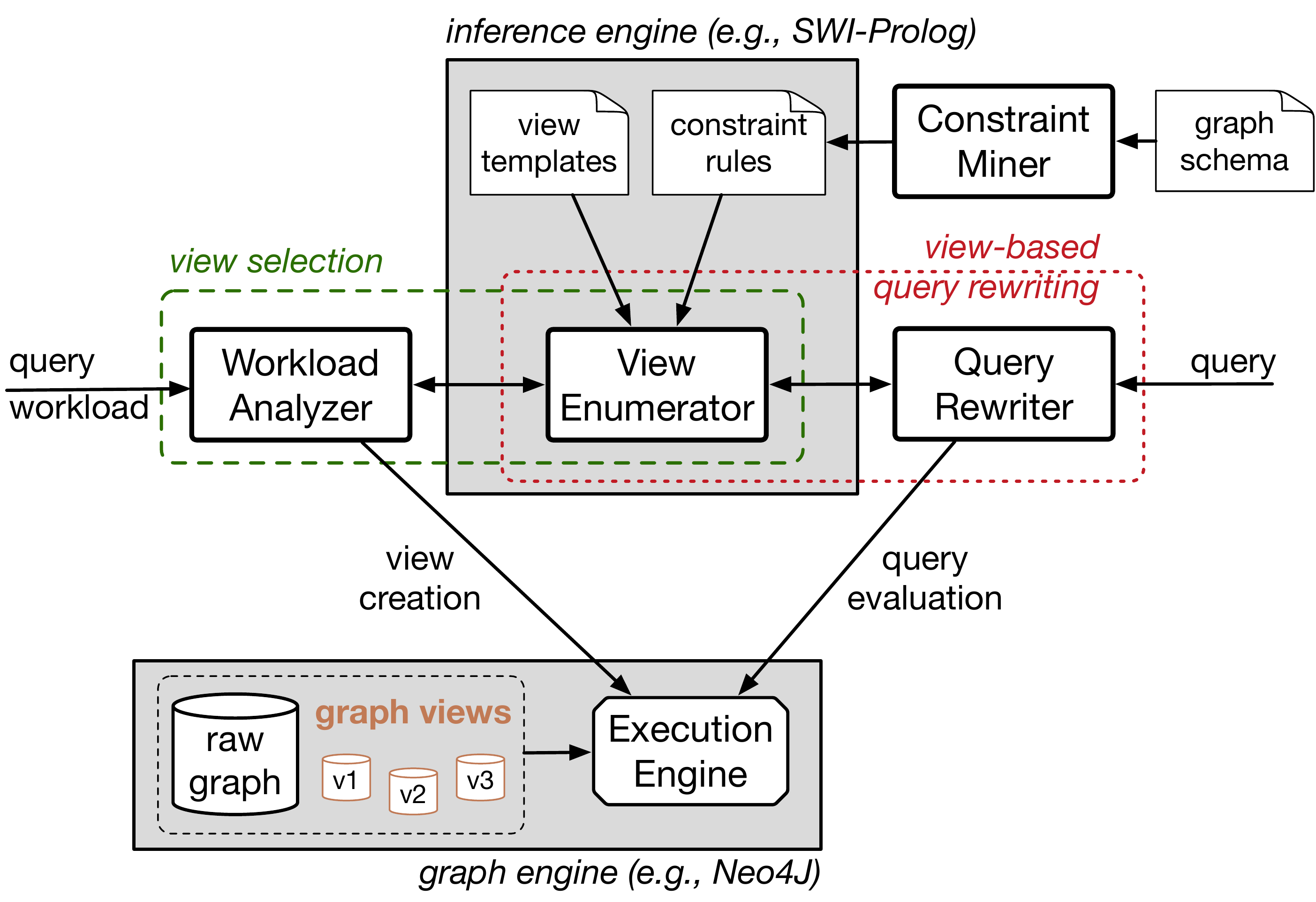}
\end{center}
\caption{Architecture of \sysname.}
\label{fig:arch}
\end{figure}



\sysname's architecture is depicted in Figure~\ref{fig:arch}.  Users submit queries in a language that includes graph pattern constructs expressed in Cypher~\cite{cypher,Francis2018} and relational constructs expressed in SQL. They use the former to express path traversals, and the latter for filtering and aggregation operations. This query language, described in \autoref{s:prelims}, is capable of capturing many of the applications described above.

\sysname supports two main view-based operations: (i)~\emph{view selection}, i.e., given a set of queries, identify the most useful views to materialize for speeding up query evaluation, accounting for a space budget and various cost components; and (ii)~\emph{view-based query rewriting}, i.e., given a submitted query, determine how it can be rewritten given the currently materialized views in the system to improve the query's execution time by leveraging the views. These operations are detailed in \autoref{s:viewops}. The \emph{workload analyzer} drives view selection, whereas the \emph{query rewriter} is responsible for the view-based query rewriting.

An essential component in both of these operations is the \emph{view enumerator}, which takes as input a query and a graph schema, and produces candidate views for that query. A subset of these candidates will be selected for materialization during view selection and for rewriting a query during view-based query rewriting. 
As we show in \autoref{s:venum}, \sysname follows a novel constraint-based view enumeration approach. In particular, \sysname's {\em constraint miner} extracts (explicit) constraints directly present in the schema and queries, and uses constraint mining rules to derive further (implicit) constraints.
\sysname employs an inference engine (we use Prolog in our implementation) to perform the actual view enumeration, using a set of view templates it expresses as inference rules. Further, it injects the mined constraints during enumeration, leading to a significant reduction in the space of candidate views.
Moreover, this approach allows us to easily include new view templates, extending the capabilities of the system, and alleviates the need for writing complicated code to perform the enumeration.

\sysname uses an \emph{execution engine} component to create the views that are output by the workload analyzer, and to evaluate the rewritten query output by the query rewriter. In this work, we use Neo4j's execution engine~\cite{neo4j} for the storage of materialized views, and to execute graph pattern matching queries. However, our query rewriting techniques can be applied to other graph query execution engines, so long as their query language supports graph pattern matching clauses.
\section{Preliminaries}
\label{s:prelims}

\subsection{Graph Data Model}
\label{ss:dm}


We adopt property graphs as our data model~\cite{PGmodel}, in which both vertices and edges are typed and may have properties in the form of key-value pairs. This {\it schema} captures constraints such as domain and range of edge types. In our provenance graph example of \autoref{sec:example}, an edge of type ``read'' only connects vertices of type ``job'' to vertices of type ``file'' (and thus never connects two vertices of type ``file'').  
As we show later, such schema constraints play an essential role in view enumeration (\autoref{s:venum}). Most mainstream graph engines~\cite{neo4j, graphx, titan, gremlin, sqlserver_graph, oracle_graph, agensgraph} provide support for this data model (including the use of schema constraints). 


\subsection{Query Language}
\label{ss:ql}

To address our query language requirements discussed in \autoref{s:overview}, \sysname combines regular path queries with relational constructs in its query language. In particular, it leverages the graph pattern specification from Neo4j's Cypher query language~\cite{cypher} and combines it with relational constructs for filters and aggregates. This hybrid query language resembles that of recent industry offerings for graph-structured data analytics, such as those from AgensGraph DB~\cite{agensgraph}, SQL Server~\cite{sqlserver_graph}, and Oracle's PGQL~\cite{pgql}.

As an example, consider the job blast radius use case introduced in \autoref{s:intro} (\autoref{fig:blast_radius}). \autoref{listing:blast_radius} illustrates a query that combines OLAP and anchored path constructs to rank jobs in its blast radius based on average CPU consumption.

\begin{listing}[htb!]
\sqlcode{listings/blast_radius.sql}
\caption{Job blast radius query over raw graph.}
\label{listing:blast_radius}
\end{listing}

Specifically, the query in~\autoref{listing:blast_radius} ranks jobs up to $10$ hops away in the downstream of a job (\texttt{q\_j1}). In the example query, this is accomplished in Cypher syntax by using a {\em variable length path} construct of up to $8$ hops (\texttt{-[r*0..8]->}) between two file vertices (\texttt{q\_f1} and \texttt{q\_f2}), where the two files are endpoints of the first and last edge on the complete path, respectively.

Although we use the Cypher syntax for our graph queries, our techniques can be coupled with any query language, as long as it can express graph pattern matching and schema constructs.


\subsection{Graph Views}
\label{s:graph-views-overview}


We define a {\em graph view} over a graph $G$ as the graph query $Q$ to be executed against $G$. This definition is similar to the one first introduced by Zhuge and Garcia-Molina~\cite{graphviews}, but extended to allow the results of $Q$ to also contain new vertices and edges in addition to those in $G$. A {\em materialized graph view} is a physical data object containing the results of executing $Q$ over $G$.

\sysname can support a wide range of graph views through the use of {\em view templates}, which are essentially inference rules, as we show in \autoref{s:venum}. Among all possible views, we identify two classes, namely \emph{\vtype{spanners}} and \emph{\vtype{sparsifiers}}, which are sufficient to capture most of the use cases that we have discussed so far. Intuitively, \vtype{spanners} result from operations over paths (i.e., path contractions), whereas \vtype{sparsifiers} are obtained via summarization operations (filters or aggregates) that reduce the number of edges and/or vertices in the original graph.

\begin{figure}[htb!]
\centering\scalebox{0.9}{\includegraphics[width=\columnwidth]{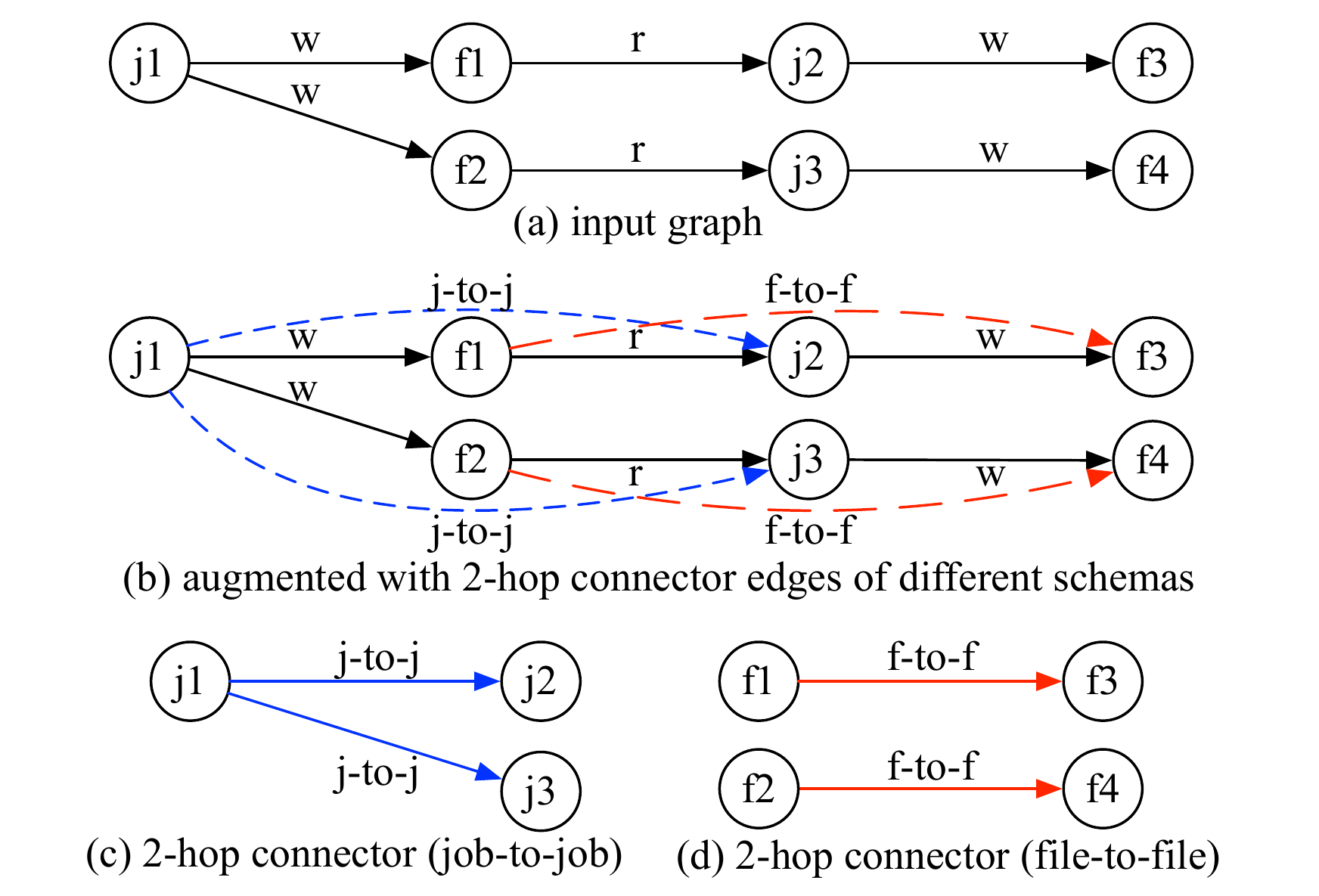}}
\caption{Construction of different $2$-hop \protect\vtype{spanner} graph views over a heterogeneous network (namely, a data lineage graph) with two vertex types ($N=2$) and two edge types ($M=2$).}
\label{fig:spanner_view}
\end{figure}

As an example, \autoref{fig:spanner_view} shows the construction of $2$-hop same-vertex-type \vtype{spanner} views over a data lineage graph (similar to the graph of \autoref{fig:blast_radius}). In \autoref{fig:spanner_view}(a), the input graph contains two types of vertices, namely jobs (vertex labels with a $j$ prefix) and files (vertex labels with an $f$ prefix), as well as two types of edges, namely $w$ (a job {\em writes to} a file) and $r$ (a file {\em is read by} a job). \autoref{fig:spanner_view}(b) shows two different types of \vtype{spanner} edges: the first contracts $2$-hop paths between pairs of job vertices (depicted as blue dashed edges); the second contracts $2$-hop paths between pairs of file vertices (depicted as red dashed edges). Finally, \autoref{fig:spanner_view}(c) shows the two resulting \vtype{spanner} graph views: one containing only the {\em job-to-job} \vtype{spanner} edges (left), and one with only the {\em file-to-file} \vtype{spanner} edges (right). 

We defer the formal definition and more examples of \vtype{spanners} and \vtype{sparsifiers} to \autoref{s:graph-views-detailed}, given that we consider their definition as a means to an end rather than a fundamental contribution of this work. Here and in Sections~\ref{s:venum} and~\ref{s:viewops}, we only provide sufficient information over graph views for the reader to follow our techniques. As path operations tend to be the most expensive operations in graphs, our description will pivot mostly around \vtype{spanners}. Materializing such expensive operations can lead to significant performance improvements at query evaluation time.  Moreover, the semantics of \vtype{spanners} is less straightforward than that of \vtype{sparsifiers}, which resemble their relational counterparts (filters and aggregates). Note, however, that the techniques described here are generic and apply to any graph views that can are expressible as view templates.


\section{Constraint-Based View Enumeration}
\label{s:venum}

Having introduced a preliminary set of graph view types in \autoref{s:graph-views-overview} (see also \autoref{s:graph-views-detailed} for more examples), we now describe \sysname's view enumeration process, which generates candidate views for the graph expressions of a query. As discussed in \autoref{s:overview}, view enumeration is central to \sysname (see \autoref{fig:arch}), as it is used in both view selection and query rewriting, which are described in \autoref{s:viewops}. Given that the space of view candidates can be considerable, and that view enumeration is on the critical path of view selection and query rewriting, its efficiency is crucial.

\begin{figure}[t!]
\centering\includegraphics[width=\columnwidth]{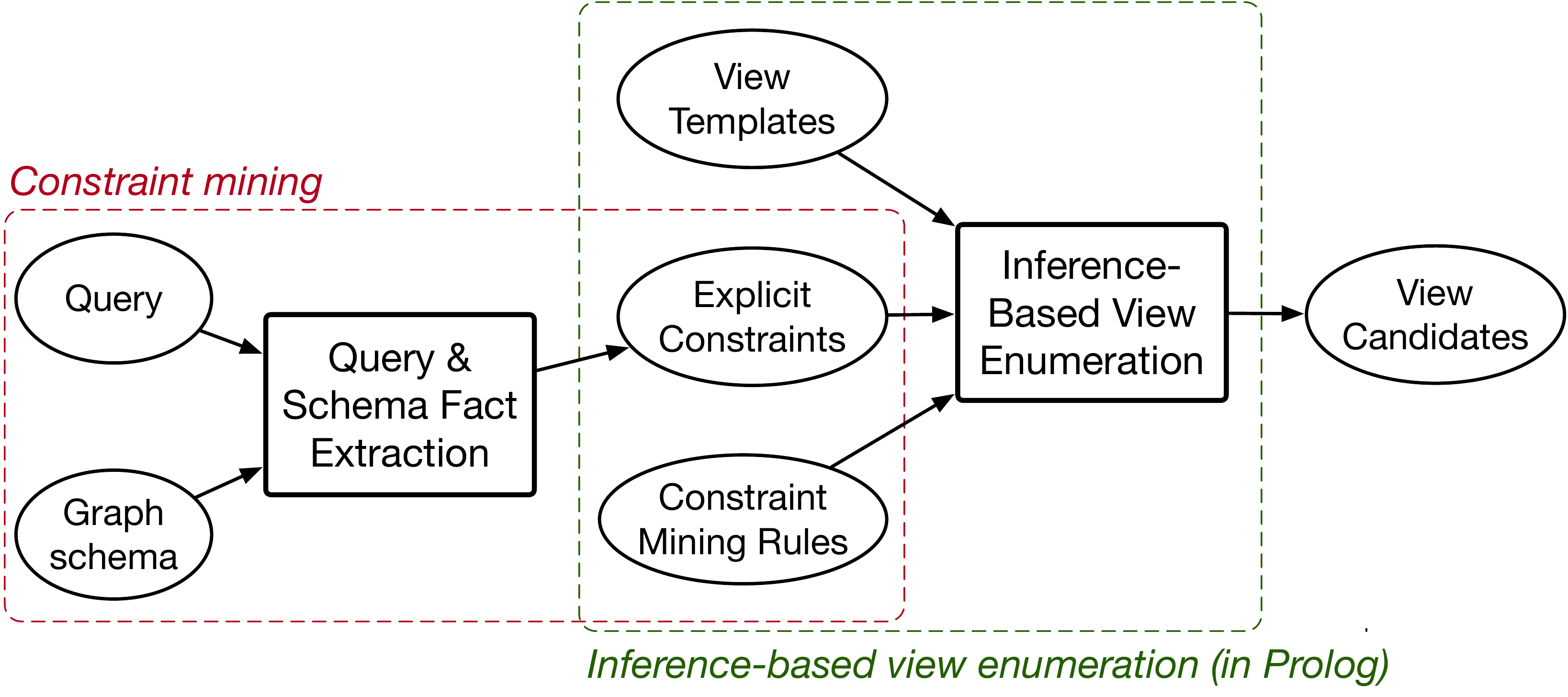}
\caption{Constraint-based view enumeration in \sysname.}
\label{fig:view_enumeration}
\end{figure}

In \sysname we use a novel approach, which we call \emph{constraint-based view enumeration}. This approach mines constraints from the query and the graph schema and injects them at view enumeration time to drastically reduce the search space of graph view candidates. \autoref{fig:view_enumeration} depicts an overview of our approach. Our view enumeration takes as input a query, a graph schema and a set of declarative \emph{view templates}, and searches for instantiations of the view templates that apply to the query. 
\sysname expresses view templates as inference rules, and employs an inference engine (namely, Prolog)\footnote{While a subset of \sysname's constraint mining rules are expressible in Datalog, Prolog provides native support (i.e., without extensions) for aggregation and negation, and we use both in \vtype{sparsifier} view templates. \sysname also relies on higher-order predicates (e.g., \texttt{setof}, \texttt{findall}) in constraint mining rules, which Datalog does not support.} to perform view enumeration through rule evaluation.
Importantly, \sysname generates both \emph{explicit} constraints, extracted directly from the query and schema, and \emph{implicit} ones, generated via constraint mining rules. This constraint mining process identifies structural properties from the schema and query that allow it to significantly prune the search space for view enumeration, discarding infeasible candidates (e.g., job-to-job edges or $3$-hop \vtype{spanners} in our provenance example).

Apart from effectively pruning the search space of candidate views, \sysname's view enumeration provides the added benefit of not requiring the implementation of complex transformations and search algorithms---the search is performed by the inference engine automatically via our view templates and efficiently via the constraints that we mine. Moreover, view templates are readily extensible to modify the supported set of views. In our current implementation, we employ SWI-PL's inference engine~\cite{swipl}, which gives us the flexibility to seamlessly add new view templates and constraint mining rules via Prolog rules. On the contrary, existing techniques for view enumeration in the relational setting typically decompose a query through a set of application-specific transformation rules~\cite{mqoSellis} or by using the query optimizer~\cite{volcanoMQO}, and then implement search strategies to navigate through the candidate views. Compared to our constraint-based view enumeration, these approaches require higher implementation effort and are inflexible when it comes to adding or modifying complex transformation rules.

We detail the two main parts of our constraint-based view enumeration, namely the constraint mining and the inference-based view enumeration, in Sections~\ref{ss:strschema} and~\ref{ss:venumalgo}, respectively.

\subsection{Mining Structural Graph Constraints}
\label{ss:strschema}

\sysname exploits information from the graph's schema and the given query in the form of constraints to prune the set of candidate views to be considered during view enumeration. It mines two types of constraints:
\begin{itemize}
  \item \textbf{Explicit constraints} (\autoref{ss:explicit-constr}) are first-order logic statements (facts) extracted from the schema and query (e.g., that files do not write files, only jobs);
  \item \textbf{Implicit constraints} (\autoref{ss:implicit-constr}) are not present in the schema or query, but are derived by combining the explicit constraint facts with constraint mining rules. These constraints are essential in \sysname, as they capture structural properties that otherwise cannot be inferred by simply looking at the input query or schema properties.
\end{itemize}

\subsubsection{Extracting explicit constraints}
\label{ss:explicit-constr}

The first step in our constraint-based view enumeration is to extract explicit constraints (facts) from the query and schema.

\mypara{Transforming the query to facts.} 
Our view enumeration algorithm goes over the query's \texttt{MATCH} clause, i.e., its graph pattern matching clause, and for each vertex and edge in the graph pattern, \sysname's constraint miner emits a set of Prolog facts. In our running example of the job ``blast radius'' query (\autoref{listing:blast_radius}), \sysname extracts the following facts from the query:


\scriptsize
\begin{verbatim}
queryVertex(q_f1). queryVertex(q_f2).
queryVertex(q_j1). queryVertex(q_j2).
queryVertexType(q_f1, 'File').
queryVertexType(q_f2, 'File').
queryVertexType(q_j1, 'Job').
queryVertexType(q_j2, 'Job').
queryEdge(q_j1, q_f1). queryEdge(q_f2, q_j2).
queryEdgeType(q_j1, q_f1, 'WRITES_TO').
queryEdgeType(q_f2, q_j2, 'IS_READ_BY').
queryVariableLengthPath(q_f1, q_f2, 0, 8).
\end{verbatim}
\normalsize

The above set of facts contains all named vertices and edges in the query's graph pattern, along with their types, and any variable-length regular path expression (\texttt{queryVariableLengthPath( X,Y,L,U)} corresponds to a path between nodes \texttt{X} and \texttt{Y} of length between \texttt{L} and \texttt{U}). In~\autoref{listing:blast_radius}, a variable-length regular path of up to 8 hops is specified between query vertices \texttt{q\_f1} and \texttt{q\_f2}.

\mypara{Transforming the schema to facts.} 
Similar to the extraction of query facts, our view enumeration algorithm goes over the provided graph schema and emits the corresponding Prolog rules. For our running example of the data lineage graph, there are two types of vertices (files and jobs) and two types of edges representing the producer-consumer data lineage relationship between them.  Hence, the set of facts extracted about this schema is:

\scriptsize
\begin{verbatim}
schemaVertex('Job'). schemaVertex('File').
schemaEdge('Job', 'File', 'WRITES_TO').
schemaEdge('File', 'Job', 'IS_READ_BY').
\end{verbatim}
\normalsize

\subsubsection{Mining implicit constraints}
\label{ss:implicit-constr}

Although the explicit query and schema constraints that we introduced so far restrict the view enumeration to consider only views with meaningful vertices and edges (i.e., that appear in the query and schema), they still allow many infeasible views to be considered. For example, if we were to enumerate $k$-hop \vtype{spanners} between two files to match the variable-length path in the query of \autoref{listing:blast_radius}, all values of $k \ge 2$ would have to be considered.  However, given the example schema, we know that only even values of $k$ are feasible views. Similarly, since the query specifies an upper limit $u=8$ in the number hops in the variable-length path, we should not be enumerating $k$-hop \vtype{spanners} with $k \ge 8$.

To derive such implicit schema and query constraints that will allow us to significantly prune the search space of views to consider, \sysname uses a library of \emph{constraint mining rules}\footnote{The collection of constraint mining rules \sysname provides is readily extensible, and users can supply additional constraint mining rules if desired.} for the schema and query. \autoref{listing:structural_properties} shows an example of such a constraint mining rule for the schema. Rule \texttt{schemaKHopPath}, expressed in Prolog in the listing, infers all valid $k$-hop paths given the input schema. Note that the rule takes advantage of the explicit schema constraint \texttt{schemaEdge} to derive this more complex constraint from the schema, and it considers two instances of \texttt{schemaKHopPath} different if and only if all values in the resulting unification tuple are different. For example, a \texttt{schemaKHopPath(X=`Job',Y=`Job',K=2)} unification, i.e., a \emph{job-to-job} 2-hop \vtype{spanner} is different from a \texttt{schemaKHopPath(X=`File',Y=`File',K=2)}, i.e., a \emph{file-to-file} 2-hop \vtype{spanner}. For completeness, we also provide a procedural version of this constraint mining rule  in \autoref{ss:additional-constraint-mining-rules} (\autoref{algo:constraint_mining_example}). When contrasted with the Prolog rule, the procedural version is not only more complex, but it also explores a larger search space. This is because the procedural function cannot be injected at view enumeration time as an additional rule together with other inference rules that further bound the search space (see \autoref{ss:venumalgo}).

\begin{listing}[htb!]
\prologcode{listings/structural_properties.pl}
\caption{Example of constraint mining rule for the graph schema.}
\label{listing:structural_properties}
\end{listing}

Similar constraint mining rules can be defined for the query. \autoref{ss:additional-constraint-mining-rules} (\autoref{listing:extraction_rules_a}) shows examples of such rules, e.g., to bound the length of considered  $k$-hop \vtype{spanners} (in case such limits are present in the query's graph pattern), or to ensure that a node is the source or sink in a \vtype{spanner}.

These constraints play a crucial role in limiting the search space for valid views.  As we show in \autoref{ss:venumalgo}, by injecting such rules at view enumeration time,  \sysname can automatically derive implicit knowledge from the schema and query to significantly reduce the search space of considered views. Interestingly, \sysname can derive this knowledge only on demand, in that the constraint mining rules get fired only when required and do not blindly derive all possible facts from the query and schema. 

Importantly, the combination of both schema and query constraints is what makes it possible for our view enumeration approach to reducing the search space of possible rewritings.  As an example, consider the process of enumerating valid $k$-hop \vtype{spanner} views. 
Without any query constraints, the number of such views equals the number of  $k$-length paths over the {\em schema graph}, which has $M$ {\em schema edges}. 
While there exists no closed formula for this combination, when the schema graph has one or more cycles (e.g., a schema edge that connects a schema vertex to itself), at least $M^k$ $k$-hop schema paths can be enumerated. This space is what the \texttt{schemaKHopPath} schema constraint mining rule would search over, were it not used in conjunction with query constraint mining rules on view template definitions. \sysname's {\em constraint-based view inference} algorithm enumerates a significantly smaller number of views for input queries, as the additional constraints enable its inference engine to efficiently search by early-stopping on branches that do not yield feasible rewritings.


\subsection{Inference-based View Enumeration}
\label{ss:venumalgo}

\begin{listing}[t!]
\prologcode{listings/spanner_views.pl}
\caption{Example view template definitions for \protect\vtype{spanners}.}
\label{listing:spanner_templates}
\end{listing}

As shown in \autoref{fig:view_enumeration}, \sysname's view enumeration takes as input (i)~a query, (ii)~a graph schema, and (iii)~a set of view templates. As described in \autoref{ss:strschema}, the query and schema are used to mine explicit and implicit constraints. Both the view templates and the mined constraints are rules that are passed to an inference engine to generate candidate views via rule evaluation. Hence, we call this process \emph{inference-based view enumeration}. The view templates drive the search for candidate views, whereas the constraints restrict the search space of considered views. This process outputs a collection of instantiated candidate view templates, which are later used either by the \emph{workload analyzer} module (see \autoref{fig:arch}) when determining which views to materialize during view selection (\autoref{ss:viewsel}), or by the \emph{query rewriter} to rewrite the query using the materialized view (\autoref{aquv}) that will lead to its most efficient evaluation.

\autoref{listing:spanner_templates} shows examples of template definitions for \vtype{spanner} views. Each view template is defined as a Prolog rule and corresponds to a type of \vtype{spanner} view. For instance, \texttt{kHopConnect\-or(X,Y,XTYPE,YTYPE,K)} corresponds to a \vtype{spanner} of length \texttt{K} between nodes \texttt{X} and \texttt{Y}, with types \texttt{XTYPE} and \texttt{YTYPE}, respectively. An example instantiation of this template is \texttt{kHopConnect\-or(X,Y,`Job',`Job',2)}, which corresponds to a concrete {\em job-to-job} 2-hop \vtype{spanner} view. This view can be translated to an equivalent Cypher query, which will be used either to materialize the view or to rewrite the query using this view, as we explain in \autoref{s:viewops}. Other templates in the listing are used to produce \vtype{spanners} between nodes of the same type or source-to-sink \vtype{spanners}. Additional view templates can be defined in a similar fashion, such as the ones for \vtype{sparsifier} views that we provide in \autoref{ss:additional-view-templates} (\autoref{listing:extraction_rules_b}).

Note that the body of the view template is defined using the query and schema constraints, either the explicit ones (e.g., \texttt{query\-VertexType}) coming directly from the query or schema, or via the constraint mining rules (e.g., \texttt{queryKHopPath}, \texttt{schema\-KHopPath}), as discussed in
\autoref{ss:strschema}. For example, \texttt{kHop\-Connector}'s body includes two explicit constraints to check the type of the two nodes participating in the \vtype{spanner}, a query constraint mining rule that will check whether there is a valid $k$-hop path between the two nodes in the query, and a schema constraint mining rule that will do the same check on the schema.




\KK{Given that view enumeration has become a very central part of the paper, I am wondering if we can devise an experiment to show how Prolog is better in the states considered than a procedural version, how you can screw up with wrong clause order in the body, etc.}

Having gathered all query and schema facts, \sysname's view enumeration algorithm performs the actual view candidate generation.  In particular, it calls the inference engine for every view template. As an example, assuming an upper bound of $k=10$ --- in ~\autoref{listing:blast_radius} we have a variable-length path of at most 8 hops between 2 {\em File} vertices, and each of these two vertices is an endpoint for another edge --- the following are valid instantiations of the \texttt{kHopConnector(X,Y,XTYPE,YTYPE,K)} view template for query vertices \texttt{q\_j1} and \texttt{q\_j2} (the only vertices projected out of the \texttt{MATCH} clause):

\scriptsize
\begin{verbatim}
(X='q_j1', Y='q_j2', XTYPE='Job', XTYPE='Job', K=2)
(X='q_j1', Y='q_j2', XTYPE='Job', XTYPE='Job', K=4)
(X='q_j1', Y='q_j2', XTYPE='Job', XTYPE='Job', K=6)
(X='q_j1', Y='q_j2', XTYPE='Job', XTYPE='Job', K=8)
(X='q_j1', Y='q_j2', XTYPE='Job', XTYPE='Job', K=10)
\end{verbatim}
\normalsize

Or, in other words, the unification \texttt{(X=`q\_j1', Y=`q\_j2', XTYPE=`Job', XTYPE=`Job', K=2)} for the view template \texttt{kHopConnector(X,Y,XTYPE,\-YTYPE,K)} implies that a view where each edge contracts a path of length $k=2$ between two nodes of type {\em Job} is feasible for the query in~\autoref{listing:blast_radius}. \joana{add unifications for intermediate rules if we think we need that level of detail}

Similarly, \sysname generates candidates for the remaining templates of Listings~\ref{listing:spanner_templates} and~\ref{listing:extraction_rules_b}. 
For each candidate it generates, \sysname's inference engine also outputs a rewriting of the given query that uses this candidate view, which is crucial for the view-based query rewriting, as we show in \autoref{aquv}. 
Finally, each candidate view incurs different costs (see \autoref{ss:costmodel}), and not every view is necessarily materialized or chosen for a rewrite.


\section{View Operations}
\label{s:viewops}

In this section, we present the main two operations that \sysname supports:
selecting views for materialization given a set of queries (\autoref{ss:viewsel}) and view-based rewriting of a query
given a set of pre-materialized views (\autoref{aquv}).
To do this, \sysname uses a cost model to estimate the size and cost of views, which we describe next (\autoref{ss:costmodel}).


\subsection{View Size Estimation and Cost Model}
\label{ss:costmodel}


While some techniques for relational cost-based query optimization may at times
target filters and aggregates, most efforts in this area have primarily focused on
join cardinality estimation~\cite{QueryProcessingSurvey}. This is in part
because joins tend to dominate query costs, but also because estimating join
cardinalities is a harder problem than, e.g., estimating the cardinality of filters.
Furthermore, \sysname can leverage existing techniques for cardinality estimation of filters and aggregates in relational query optimization for
\vtype{sparsifiers}.

Therefore, here we detail our cost model contribution as it relates to \vtype{spanner} views, which can be seen as the graph counterpart of relational joins.
We first describe how we estimate the size of \vtype{spanner} views, which we then use to define our cost model.

\mypara{Graph data properties} During initial data loading and subsequent data updates, \sysname maintains the following graph properties: (i)~\emph{vertex cardinality} for each vertex type of the raw graph; and (ii)~\emph{coarse-grained out-degree distribution summary statistics}, i.e., the 50th, 90th, and 95th out-degree for each vertex type of the raw graph. \sysname uses these properties to estimate the size of a view.

\mypara{View size estimation.} Estimating the size of a view is essential for the \sysname's view operations.
First, it allows us to determine if a view will fit in a space budget during view selection. Moreover, our cost components (see below) use view size estimates to assess the benefit of a view both in view selection and view-based query rewriting.

In \sysname, we estimate the size of a view as the number of edges that it has when materialized since the number of edges usually dominates the number of vertices in real-world graphs. 
Next, we observe that the number of edges in a $k$-hop \vtype{spanner} over a graph $G$ equals the number of $k$-length simple paths in $G$. We considered using the following estimator for Erdos-Renyi random graphs~\cite{bollobas01}, where edges are generated uniformly at random, and the expected number of $k$-length simple paths can be computed as:

\vspace{-2pt}
\begin{equation}
  \hat{E}(G, k) = {n \choose k+1} \cdot \bigg[\frac{m}{{n \choose 2}}\bigg]^k
\end{equation}
\vspace{-2pt}

\noindent where $n = |V|$ is the number of vertices in $G$, and $m = |E|$ is the number of edges in $G$.
However, we found that for real-world graphs, the independence assumption does not hold for vertex connectivity, as the degree distributions are not necessarily uniform and edges are usually correlated.
As a result, this formula significantly underestimates --- by several orders of magnitude --- the number of directed $k$-length paths in real-world graphs.
Therefore, for a directed graph $G$ that is \emph{homogeneous} (i.e., has only one type of vertex, and one type of edge), we define the following estimator for its number of $k$-length paths:

\vspace{-2pt}
\begin{equation}
  \hat{E}(G, k, \alpha) = n \cdot deg_{\alpha}^k
\end{equation}
\vspace{-2pt}

\noindent where $n = |V|$ is the number of vertices in $G$, and $deg_{\alpha}$ is the 
  $\alpha$-th percentile out-degree of vertices in $G$ ($0<\alpha\le 100)$.

For a directed graph $G$ that is \emph{heterogeneous} (i.e., has more than one
type of vertex and/or more than one type of edge), an estimator for its number of $k$-length paths is as follows:

\vspace{-2pt}
\begin{equation}
  \hat{E}(G, k, \alpha) = \sum_{t\in T_G} n_t \cdot (deg_{\alpha}(n_{i}))^k
\end{equation}
\vspace{-2pt}

\noindent where $T_{G}$ is the set of types of vertices in $G$ that are edge sources (i.e., are the domain of at least one type of edge), $n_{t}$ is the number of vertices of type $t \in T_{G}$, and $deg_{\alpha}(n_{t})$ is the maximum out-degree of vertices of type $t \in T_{G}$.


Observe that if $\alpha = 100$, then $deg_{\alpha}$ is the maximum out-degree and the estimators above are upper bounds on the number of $k$-length paths in the graph. This is because there are $n$ possible starting vertices, each vertex in the path has at most $deg_{100}$ (or $deg_{100}(n_i)$) neighbors to choose from for its successor vertex in the path, and such a choice has to be made $k$ times in a $k$-length path. 
In practice, we found that $\alpha=100$ gives a very loose upper bound, whereas $50\le\alpha\le 95$ gives a much more accurate estimate depending on the degree distribution of the graph. We present experiments on the accuracy of our estimator in~\autoref{s:experiments}.

\mypara{View creation cost.} The creation cost of a graph view refers to any computational and I/O costs incurred when computing and materializing the views' results.  Since the primitives required for computing and materializing the results of the graph views that we are interested in are relatively simple, the I/O cost dominates computational costs, and thus the latter is omitted from our cost model. Hence, the view creation cost is directly proportional to $\hat{E}(G,k,\alpha)$.



\mypara{Query evaluation cost.} The cost of evaluating a query $q$, denoted $EvalCost(q)$, is required both in the view selection and the query rewriting process. \sysname relies on an existing cost model for graph database queries as a proxy for the cost to compute a given query using the raw graph. In particular, it leverages Neo4j's~\cite{neo4j} cost-based optimizer, which establishes a reasonable ordering between all vertex scans without indexes, scans from indexes, and range scans.
As part of our future work, we plan to incorporate our findings from graph view size estimation to further improve the query evaluation cost model.

\julian{would be useful to write the equation used, which includes all the terms defined in the previous section}



\subsection{View Selection}
\label{ss:viewsel}

Given a query workload, \sysname's view selection process determines the most effective views to materialize
for answering the workload under the space budget that \sysname allocates for materializing views.
This operation is performed by the workload analyzer component, in conjunction with the view enumerator (see \autoref{s:venum}).
The goal of \sysname's view selection algorithm is to select the views that lead to the most significant performance gains relative to their cost, while respecting the space budget.

To this end, we formulate the view selection algorithm as a 0-1 knapsack problem, where the size of the knapsack is the space budget dedicated to view materialization.\footnote{The space budget is typically a percentage of the
machine's main memory size, given that we are using a main memory execution
engine. Our formulation can be easily extended to support multiple levels of memory hierarchy.}
The items that we want to fit in the knapsack are the candidate views generated
by the view enumerator (\autoref{s:venum}). The weight of each item is the view's estimated size
(see \autoref{ss:costmodel}), while the value of each item is the performance
improvement achieved by using that view divided by the view's creation
cost (to penalize views that are expensive to materialize). 
We define the performance improvement of a view $v$ for a query $q$ in the query workload $Q$ as $q$'s evaluation cost divided by the cost of evaluating the rewritten version of $q$ that uses $v$.
The performance improvement of $v$ for $Q$ is the sum of $v$'s improvement for each query in $Q$ (which is zero for the queries for which $v$ is not applicable). Note that we can extend the above formulation by adding weights to the value of each query to reflect its relative importance (e.g., based on the query's frequency to prioritize more frequent queries, or on the query's estimated execution time to prioritize more expensive queries). 

The views that the above process selects for materialization are instantiations of Prolog view templates output by the view enumeration (see \autoref{s:venum}).
\sysname's workload analyzer translates those views to Cypher and executes them against the graph to perform the actual materialization.
As a byproduct of this process, each combination of query $q$ and materialized view $v$ is accompanied by a rewriting of $q$ using $v$, which is crucial in the view-based query rewriting, described next. The rewriter component converts the rewriting in Prolog to Cypher, so that \sysname can run it on its graph execution engine.


\subsection{View-Based Query Rewriting}
\label{aquv}

Given a query and a set of materialized views, view-based rewriting is the process of finding the rewriting of the query that leads to the highest reduction of its evaluation cost by using (a subset of) the views. In \sysname, the query rewriter module (see \autoref{fig:arch}) performs this operation.


When a query $q$ arrives in the system, the query rewriter invokes the view enumerator, which generates the possible view candidates for $q$, pruning those that it has not materialized. Among the views that are output by the enumerator, the query rewriter selects the one that, when used to rewrite $q$, leads to the smallest evaluation cost for $q$.
As discussed in \autoref{s:venum}, the view enumerator outputs the rewriting of each query based on the candidate views for that query.
If this information is saved from the view selection step (which is true in our implementation), we can leverage it to choose the most efficient view-based rewriting of the query without having to invoke the view enumeration again for that query.
As we mentioned in \autoref{s:venum}, \sysname currently supports rewritings that rely on a single view. Combining multiple views in a single rewriting is left as future work.

\autoref{listing:blast_radius_2_hop} shows the rewritten version of our example query of \autoref{listing:blast_radius} that uses a 2-hop \vtype{spanner} (``job-to-job'') graph view.

\begin{listing}[h!]
\sqlcode{listings/blast_radius_2_hop.sql}
  \caption{Job blast radius query rewritten over a $2$-hop \protect\vtype{spanner} ({\em job-to-job}) graph view.}
\label{listing:blast_radius_2_hop}
\end{listing}

\section{View Definitions and Examples}
\label{s:graph-views-detailed}

Next we provide formal definitions of the two main classes of views supported in \sysname, namely \vtype{spanners} and \vtype{sparsifiers}, which were briefly described in~\autoref{s:graph-views-overview}. Moreover, we give various examples of the views from each category that are currently present in \sysname's view template library.

While the examples below are general enough to capture many different types of graph structures, they are by no means an exhaustive list of graph view templates that are {\em possible} in \sysname; as we mention in~\autoref{s:venum}, \sysname's library of view templates and constraint mining rules is readily extensible.


\subsection{Connectors}
\label{ss:spanners}

\begin{table}[t!]
\caption{\protect\vtype{Spanners} in \sysname}
\label{tab:spanners}
\begin{tabular}{p{0.25\columnwidth}p{0.65\columnwidth}}
\toprule
\textbf{Type} & \textbf{Description} \\
\midrule
Same-vertex-type \vtype{spanner} & Target vertices are all pairs of vertices with a specific vertex type.  \\
\midrule
$k$-hop \vtype{spanner} & Target vertices  are all vertex pairs that are connected through $k$-length paths. \\
\midrule
Same-edge-type \vtype{spanner}  & Target vertices are all pairs of vertices that are connected with a path consisting of edges with a specific edge type \\
\midrule
Source-to-sink \vtype{spanner}  & Target vertices are (source, sink) pairs, where sources are the vertices with no incoming edges and sinks are vertices with no outgoing edges. \\
\bottomrule
\end{tabular}
\end{table}

A \vtype{spanner} of a graph $G=(V,E)$ is a graph $G'$ such that every edge $e'=(u,v) \in E(G')$ is obtained via contraction of a single directed path between two \emph{target} vertices $u,v \in V(G)$. The vertex set $V(G')$ of the \vtype{spanner} view is the union of all target vertices with $V(G') \subseteq V(G)$. Based on this definition, a number of specialized \vtype{spanner} views can be defined, each of which differs in the target vertices that it considers. Table~\ref{tab:spanners} lists examples currently supported in \sysname.

Additionally, it is easy to compose the definition of $k$-hop \vtype{spanners} with the other \vtype{spanner} definitions, leading to more \vtype{spanner} types. As an example, the $k$-hop same-vertex-type \vtype{spanner} is a same-vertex-type \vtype{spanner} with the additional requirement that the target vertices should be connected through $k$-hop paths.

Finally, \vtype{spanners} are useful in contexts where a query's graph pattern contains relatively long paths that can be contracted without loss of generality, or when only the endpoints of the graph pattern are projected in subsequent clauses in the query.


\begin{table}[t]
\caption{\protect\vtype{Sparsifiers} in \sysname}
\label{tab:sparsifiers}
\begin{tabular}{p{0.25\columnwidth}p{0.65\columnwidth}}
\toprule
\textbf{Type} & \textbf{Description} \\
\midrule
Vertex-removal \vtype{sparsifier} & Removes vertices and connected edges that satisfy a given predicate, e.g., vertex type or vertex property. \\
\midrule
Edge-removal \vtype{sparsifier}  & Removes edges that satisfy a given predicate. \\
\midrule
Vertex-inclusion \vtype{sparsifier} & Includes vertices that satisfy the predicate, and edges where both the source and target vertices satisfy the predicate. \\
\midrule
Edge-inclusion \vtype{sparsifier} & Includes only edges that satisfy a given predicate.\\
\midrule
Vertex-aggregator \vtype{sparsifier} & Groups vertices that satisfy a given predicate, and combines them using the provided aggregate function. \\
\midrule
Edge-aggregator \vtype{sparsifier} & Groups edges that satisfy a given predicate, and combines them using the provided aggregate function. If the resulting edges do not have the same source and destination vertices, it also performs a vertex aggregation operation.\\
\midrule
Subgraph-aggregator \vtype{sparsifier} & Groups both edges and vertices that satisfy a given predicate, and combines them using the provided aggregate function.  \\
\bottomrule
\end{tabular}
\end{table}

\subsection{Summarizers}

A \vtype{sparsifier} of a graph $G=(V,E)$ is a graph $G'$ such that $V(G') \subseteq V(G)$, $E(G') \subseteq E(G)$, and at least one of the following conditions are true: (i) $|V(G')| < |V(G)|$, or (ii) $|E(G')| < |E(G)|$.  The \vtype{sparsifier} view operations that \sysname currently provides are filters that specify the type of vertices or edges that we want to preserve (inclusion filters) or remove (exclusion filters) from the original graph.\footnote{\vtype{Sparsifier} views can also include predicates on vertex/edge properties in their definitions. Using such predicates would further reduce the size of these views, but given they are straightforward, here we focus more on predicates for vertex/edge types.} Also, it provides aggregator \vtype{sparsifiers} that either group a set of vertices into a \emph{supervertex}, a group of edges into a \emph{superedge}, or a subgraph into a supervertex. Finally, aggregator \vtype{sparsifiers} require an aggregate function for each type of property present in the domain of the aggregator operation. Examples of \vtype{sparsifier} graph views currently supported are given in Table~\ref{tab:sparsifiers}. \sysname's library of template views currently does not support aggregation of vertices of different types. The library is readily extensible, however, and aggregations for vertices or edges with different types are expressible using a higher-order aggregate function to resolve conflicts between properties with the same name, or to specify the resulting aggregate type.

Lastly, \vtype{sparsifiers} can be useful when subsets of data (e.g., entire classes of vertices and edges) can be removed from the graph without incurring any side-effects (in the case of filters), or when queries refer to logical entities that correspond to groupings of one or more entity at a finer granularity (in the case of aggregators).

\section{Experimental Evaluation}
\label{s:experiments}

In this section, we experimentally confirm that by leveraging graph views (e.g., \vtype{sparsifiers} and \vtype{spanners}) we can: (i)~accurately estimate graph view sizes---\autoref{sec:exp_vs}; (ii)~effectively reduce the graph size our queries operate on---\autoref{sec:exp_sr};  and (iii)~improve query performance---\autoref{sec:exp_qr}. We start by providing details on \sysname's current implementation, and follow by introducing our datasets and query workload.


\subsection{Implementation}

We have implemented \sysname's components (see \autoref{fig:arch}) as follows. The {\em view enumerator} component (\autoref{s:venum}) uses SWI-Prolog~\cite{swipl} as its inference engine. We wrote all constraint mining rules, query constraint mining rules, and view templates using SWI-Prolog syntax, which is close to ISO-Prolog~\cite{swipl-syntax}. We wrote the knapsack problem formulation (\autoref{ss:viewsel}) in Python 2.7, using the branch-and-bound knapsack solver from Google OR tools combinatory optimization library~\cite{google-or-tools}. As shown in~\autoref{fig:arch}, \sysname uses Neo4J (version 3.2.2) for storage of raw graphs, materialized graph views, and query execution of graph pattern matching queries. All other components --- including the {\em constraint miner} --- were written in Java. 

For the experimental results below, \sysname extracted schema constraints only once for each workload, as these do not change throughout the workload. Furthermore, \sysname extracted query constraints as part of view inference only the first time we entered the query into the system. This process introduces a few milliseconds to the total query runtime --- as the number of query facts combined with schema facts is small, and in the experiments below we focus on one type of view template --- and is amortized for multiple runs of the same query.

\subsection{Datasets}

In our evaluation of different aspects of \sysname, we use a combination of publicly-available heterogeneous and homogeneous networks and a large provenance graph from \company (heterogeneous network). \autoref{tab:datasets} lists the datasets and their raw sizes. We also provide their degree distributions in \autoref{ss:degree-distributions}.

\begin{table}[htb!]
  \begin{center}
    \caption{Networks used for evaluation: \texttt{prov} and \texttt{dblp} are heterogeneous, while \texttt{roadnet-usa} and \texttt{soc-livejournal} are homogeneous and have one edge type.}
    \label{tab:datasets}
    \begin{tabular}{llrr}
      \toprule
      \textbf{Short Name} & \textbf{Type} & \textbf{$|V|$} & \textbf{$|E|$} \\
      \midrule
      prov (raw) & Data lineage & 3.2B & 16.4B \\
      prov (summarized) & Data lineage & 7M & 34M \\
      dblp-net & Publications \cite{graphdblp} & 5.1M & 24.7M \\
      soc-livejournal & Social network \cite{soc-livejournal} & 4.8M & 68.9M \\
      roadnet-usa & Road network \cite{netrepo} & 23.9M & 28.8M \\
      \bottomrule
    \end{tabular}
  \end{center}
\end{table}

For our size reduction evaluation in Section~\ref{sec:exp_sr}, we focus on gains provided by different graph views over the two heterogeneous networks: \texttt{dblp-net} and a data lineage graph from \company. The \texttt{dblp-net} graph, which is publicly available at \cite{graphdblp}, contains 5.1M vertices (authors, articles, and venues) with 24.7M with 2.2G on-disk footprint. For the second heterogeneous network, we captured a provenance graph modeling one of \company's large production clusters for a week. This \emph{raw graph} contains 3.2B vertices modeling jobs, files, machines, and tasks, and 16.4B edges representing relationships among these entities, such as job-read-file, or task-to-task data transfers. The on-disk footprint of this data is in the order of 10+ TBs.

After showing size reduction achieved by \vtype{sparsifiers} and \vtype{spanners} in Section~\ref{sec:exp_sr}, for our query runtime experiments (\autoref{sec:exp_qr}), we consider already summarized versions of our heterogeneous networks, i.e., of \texttt{dblp-net} and provenance graph. The summarized graph over \texttt{dblp-net} contains only authors and publications (``article'', ``publication'', and ``in-proc'' vertex types), totaling 3.2M vertices and 19.3M edges, which requires 1.3G on disk. The summarized graph over the raw provenance graph contains only jobs and files and their relationships, which make up its 7M vertices and 34M edges. The on-disk footprint of this data is 4.8 GBs. This allows us to compare runtimes of our queries on the Neo4j 3.2.2 graph engine, running on a 128 GB of RAM and 28 Intel Xeon 2.40GHz cores, 4 x 1TB SSD Ubuntu box. We chose to use Neo4j for storage of materialized views and query execution because it is the most widely used graph database engine as of writing, but our techniques are graph query engine-agnostic.



\subsection{Queries}
\label{sec:exp_q}

Table~\ref{tab:queries} lists the queries that we use in our evaluation of query runtimes.
Queries Q1 through Q3 are motivated by telemetry use cases at \company, and are defined as follows. The first query retrieves the job blast radius, up to 8 hops away, of all job vertices in the graph, together with their average CPU consumption property. Query Q2 retrieves the ancestors of a job (i.e., all vertices in its {\em backward data lineage}) up to 4 hops away, for all job vertices in the graph. Conversely, Q3 does the equivalent operation for {\em forward data lineage} for all vertices in the graph, also capped at 4 hops from the source vertex. Both Q2 and Q3 are also adapted for the other 3 graph datasets: on \texttt{dblp}, the source vertex type is ``author'' instead of ``job'', and on homogeneous networks \texttt{roadnet-usa} and \texttt{soc-livejournal} all vertices are included.

\begin{table}[htb!]
  \begin{center}
    \caption{Query workload.}
    \label{tab:queries}
    \begin{tabular}{lll}
      \toprule
      \textbf{Query} & \textbf{Operation} & \textbf{Result}\\
      \midrule
      Q1: Job Blast Radius & Retrieval & Subgraph \\
      Q2: Ancestors & Retrieval & Set of vertices \\
      Q3: Descendants & Retrieval & Set of vertices \\
      Q4: Path lengths & Retrieval & Bag of scalars \\
      Q5: Edge Count & Retrieval & Single scalar \\
      Q6: Vertex Count & Retrieval & Single scalar \\
      Q7: Community Detection & Update & N/A \\
      Q8: Largest Community & Retrieval & Subgraph \\
      \bottomrule
    \end{tabular}
  \end{center}
\end{table}

Next, queries Q4 through Q7 capture graph operation primitives which are commonly required for tasks in dependency driven analytics~\cite{dda-cidr-2017}.  The first, query Q4 (``path lengths''), computes a weighted distance from a source vertex to all other vertices in its forward $k$-hop neighborhood, limited to 4 hops.  It does so by first retrieving all vertices in a given vertex's 4-hop neighborhood, and then for each vertex in this result set, it performs an aggregation operation (max) over a data property (edge timestamp) of all edges in the path. Queries Q5 and Q6 both measure the overall size of the graph (edge count and vertex count, respectively).

Finally, Q7 (``community detection'') and Q8 (``largest community'') are representative of common graph analytics tasks. The former runs a 25 passes
iterative version of community detection algorithm via label-propagation, updating a {\em community} property on all vertices and edges in the graph,
while Q8 uses the community label produced by Q7 to retrieve the largest community in terms of graph size as measured by number of ``job'' vertices in each community. The label propagation algorithm used by Q7 is part of the APOC collection of graph algorithm UDFs for Neo4j~\cite{neo4j-apoc}.

For query runtimes experiments in~\autoref{sec:exp_qr}, we use the equivalent rewriting of each of these queries over a $2$-hop \vtype{spanner}. Specifically, queries Q1 through Q4 go over half of the original number of hops, and queries Q7 and Q8 run around half as many iterations of label propagation. 
These rewritings are equivalent and produce the same results as queries Q1 through Q4 over the original graph, and similar groupings of ``job'' nodes in the resulting communities. Queries Q5 and Q6 need not be modified, as they only count the number of elements in the dataset (edges and vertices, respectively).

\begin{figure*}[t]
  \begin{center}
  \includegraphics[width=0.234\linewidth]{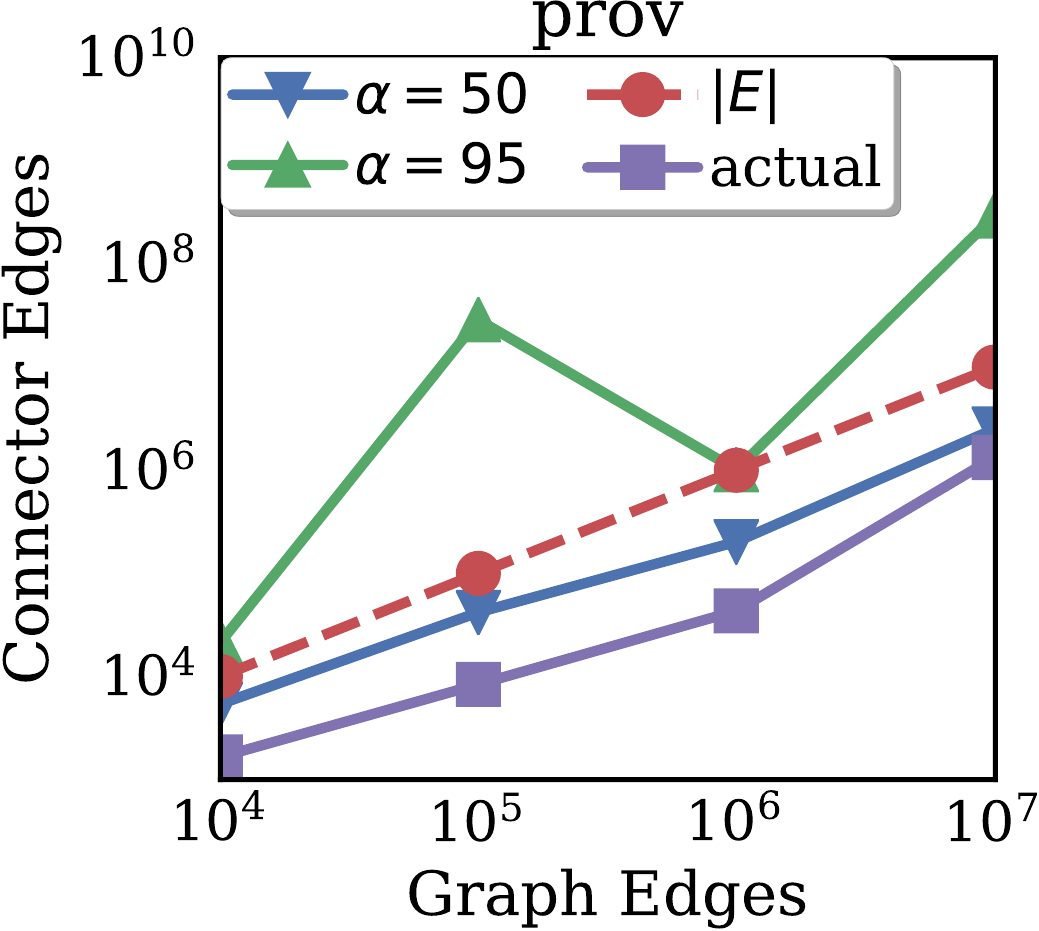}
  \includegraphics[width=0.216\linewidth]{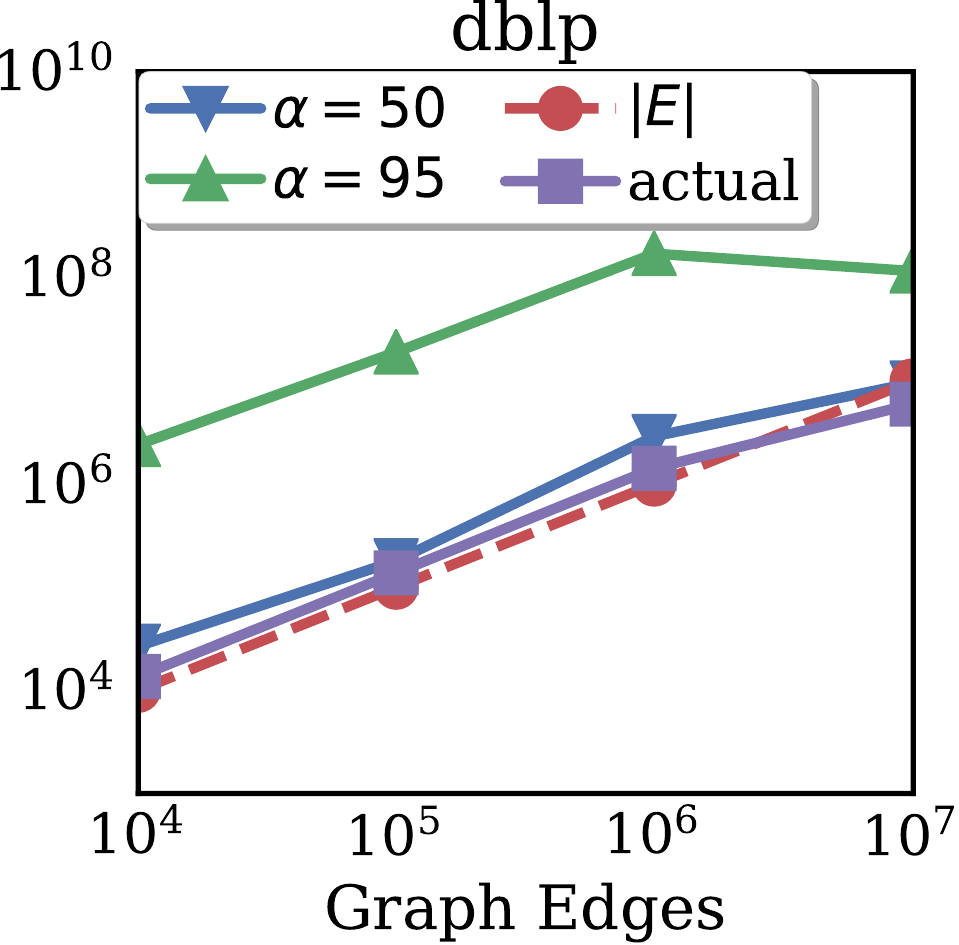}
  \includegraphics[width=0.216\linewidth]{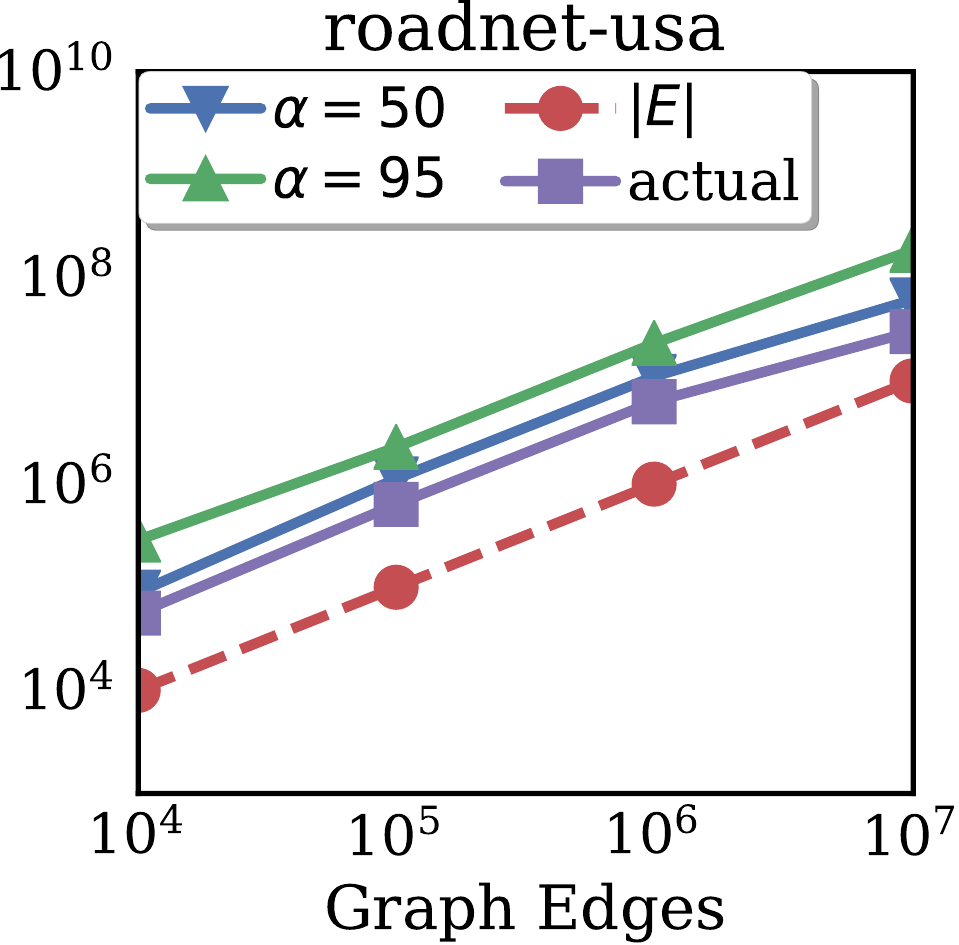}
  \includegraphics[width=0.216\linewidth]{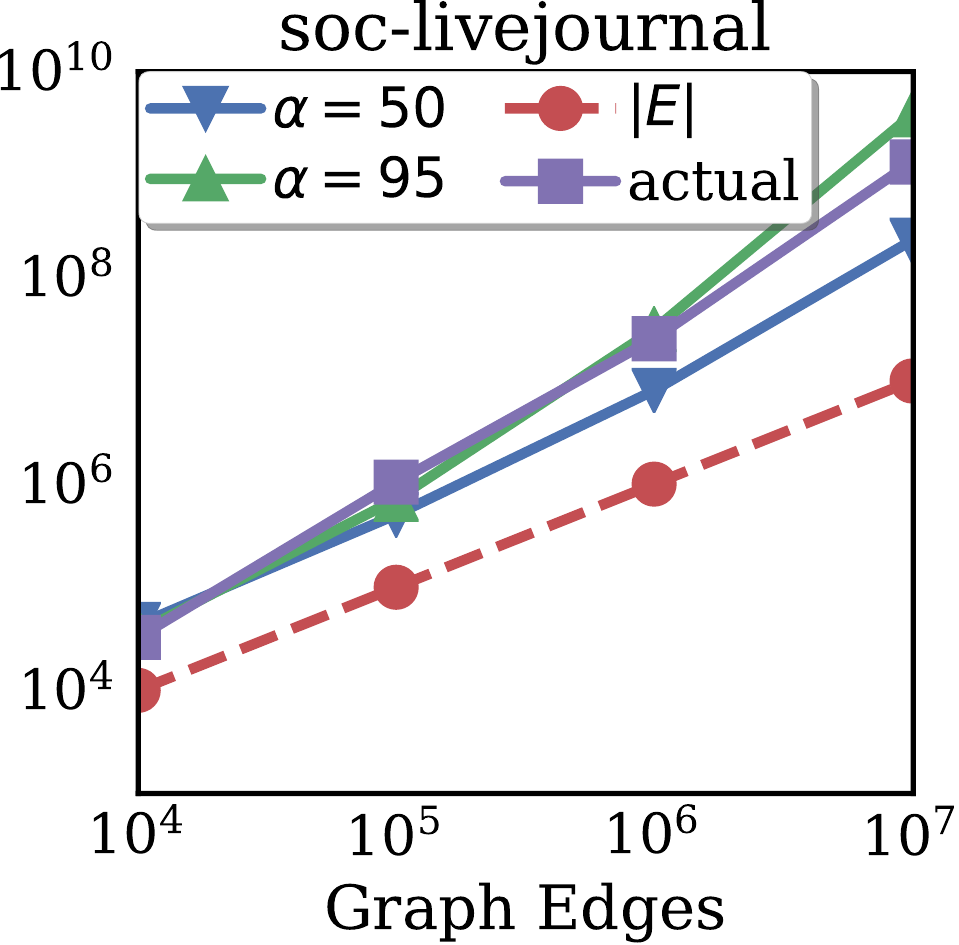}
    \caption{Estimated, actual, and original graph sizes for $2$-hop \protect\vtype{spanner} views over different datasets. Here we show estimates for two upper bound variations derived from summary statistics over the graph's degree distribution detailed in \autoref{ss:costmodel}. We also plot the original graph size ($x$-axis, and dashed $|E|$ series).  Plots are in log-log scale. \julian{what is dashed $|E|$ series?}}
  \label{fig:view_size_estimates}
  \end{center}
\end{figure*}

\subsection{View Size Estimation}
\label{sec:exp_vs}

In this section, we evaluate the accuracy of \sysname's view size estimators (\autoref{ss:costmodel}). \autoref{fig:view_size_estimates} shows our results for different heuristics estimators on the size of a $2$-hop \vtype{spanner} view materialized over the first $n$ edges of each public graph dataset.  We focus on size estimates for $2$-hop \vtype{spanners} since, similar to cardinality estimation for joins, the larger the $k$, the less accurate our estimator.  We do not report results for view size estimates for \vtype{sparsifiers}, as for these \sysname can leverage traditional cardinality estimation based on predicate selectivity for filters, as well as multi-dimensional histograms for predicate selectivity of group-by operators~\cite{QueryProcessingSurvey}.

We found that $k$-hop \vtype{spanners} in homogeneous networks are usually larger than the original graph in real-world networks. This is because $k$-length paths can exist between any two vertices in this type of a graph, as opposed to only between specific types of vertices in heterogeneous networks, such as \company's provenance graph. Note that the $\alpha = 50$ line does a good job of approximating the size of the graph as the number of edges grows. Also, for networks with a degree distribution close to a power-law, such as
\texttt{soc-livejournal}, the estimator that relies on 95th percentile out-degree ($\alpha=95$) provides an upper bound, while the one that uses the median out-degree ($\alpha=50$) of the network provides a lower bound. On other networks that do not have a power-law degree distribution, such as \texttt{road-net-usa}, the median out-degree estimator better approximates an upper bound on the size of the $k$-hop \vtype{spanner}.

In practice, \sysname relies on the estimator parameterized with $\alpha=95$ as it provides an upper bound for most real-world graphs that we have observed. Also note that the estimator with $\alpha=95$  for \texttt{prov} decreases when the original graph increases in size from $100K$ to $1M$ edges, increasing again at $10M$ edges.  This is due to a decrease in the percentage of ``job'' vertices with a large out-degree, shifting the 95th percentile to a lower value. This percentile remains the same at $10M$ edges, while the 95th out-degree for ``file'' vertices increases at both $1M$ and $10M$ edges.


\subsection{Size Reduction}
\label{sec:exp_sr}

\begin{figure}[h!]
  \begin{center}
  \includegraphics[width=0.216\textwidth]{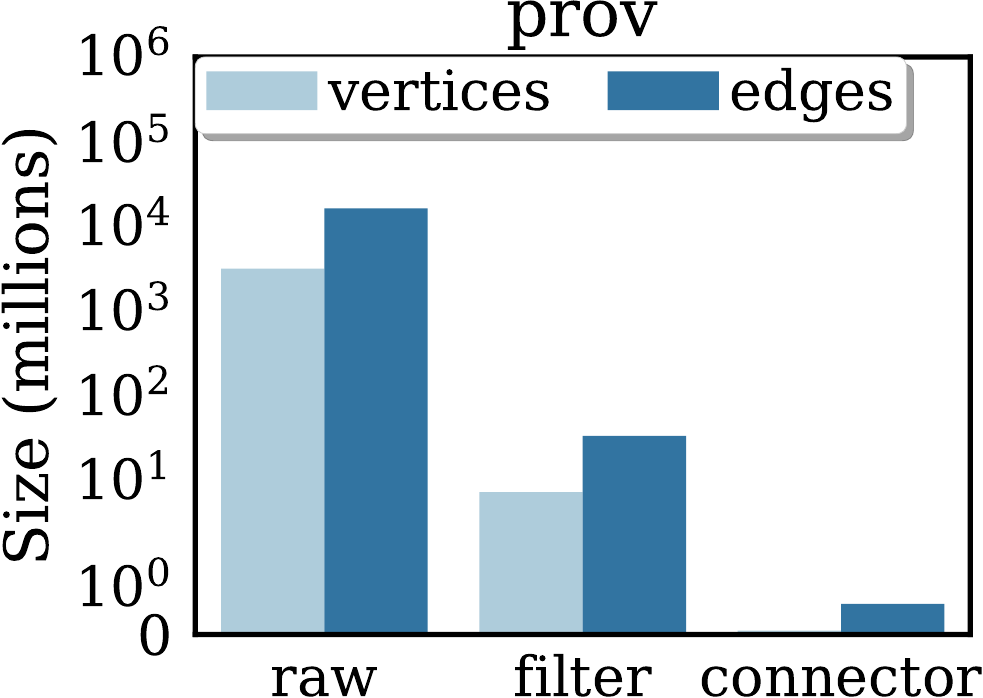}
  \hspace{1mm}
  \includegraphics[width=0.198\textwidth]{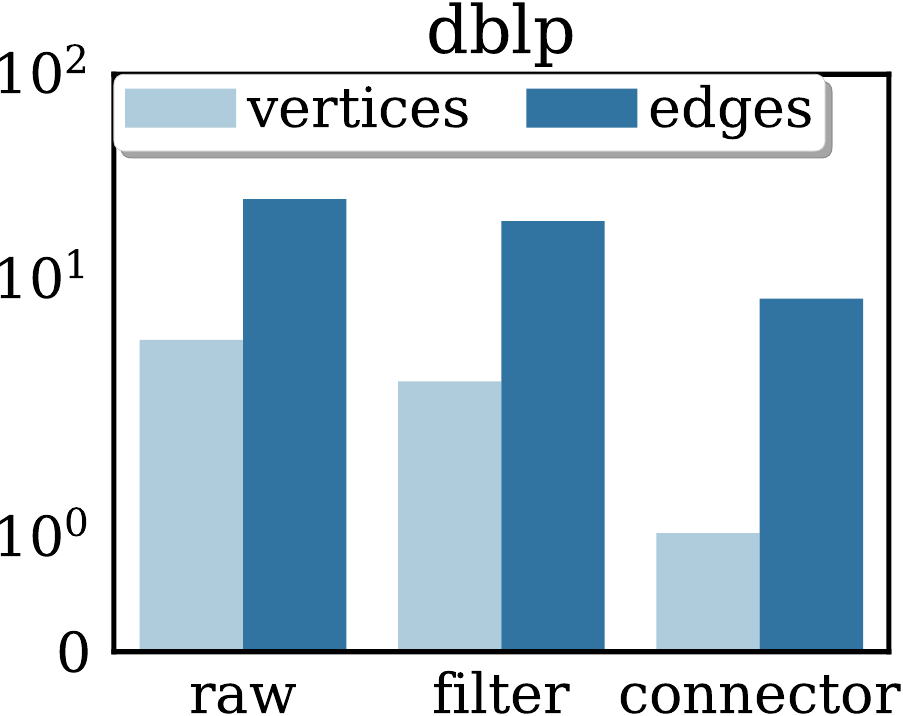}
    \caption{Effective graph size reduction when using
    \protect\vtype{sparsifier} and $2$-hop \protect\vtype{spanner} views over
    \texttt{prov} and \texttt{dblp} heterogeneous networks ($y$-axis is in log
    scale).}
 \label{fig:size_reduction}
  \end{center}
\end{figure}

This experiment shows how by applying \vtype{sparsifiers} and \vtype{spanners} over heterogeneous graphs we can reduce the effective graph size for one or more queries. Figure~\ref{fig:size_reduction} shows that for co-authorship queries over the \texttt{dblp}, and query Q1 over the provenance graph, the schema-level \vtype{sparsifier} yields up to three orders of magnitude reduction. The \vtype{spanner} yields another two orders of magnitude data reduction by summarizing the job-file-job paths in the provenance graph, and one order of magnitude by summarizing the author-publication-author paths in the \texttt{dblp} graph.

Besides the expected performance advantage since queries operate on less data, such a drastic data reduction allows us to benefit from single-machine in-memory technologies (such as Neo4j) instead of slower distributed on-disk alternatives for query evaluation, in the case of the provenance graph.
While this is practically very relevant, we do \emph{not} claim this as part of our performance advantage, and all experiments are shown as relative speed-ups against a baseline on this provenance reduced graph, with all experiments on the same underlying graph engine.

\begin{figure*}[t]
  \begin{center}
  \includegraphics[width=0.234\textwidth]{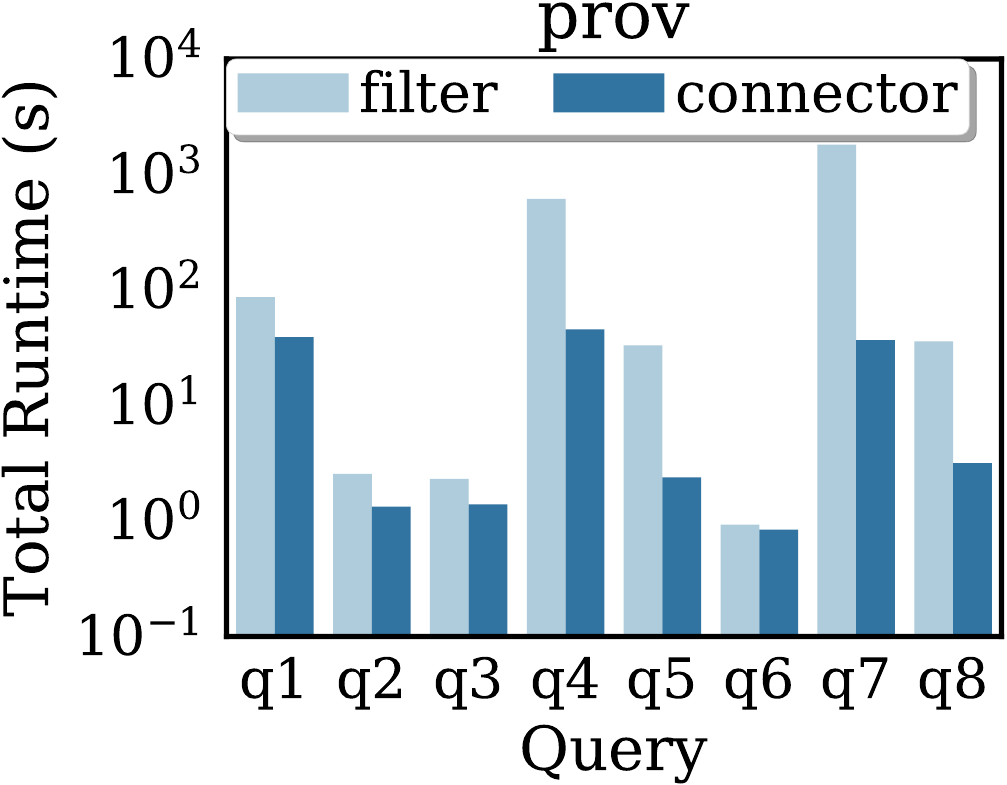}
  \includegraphics[width=0.216\textwidth]{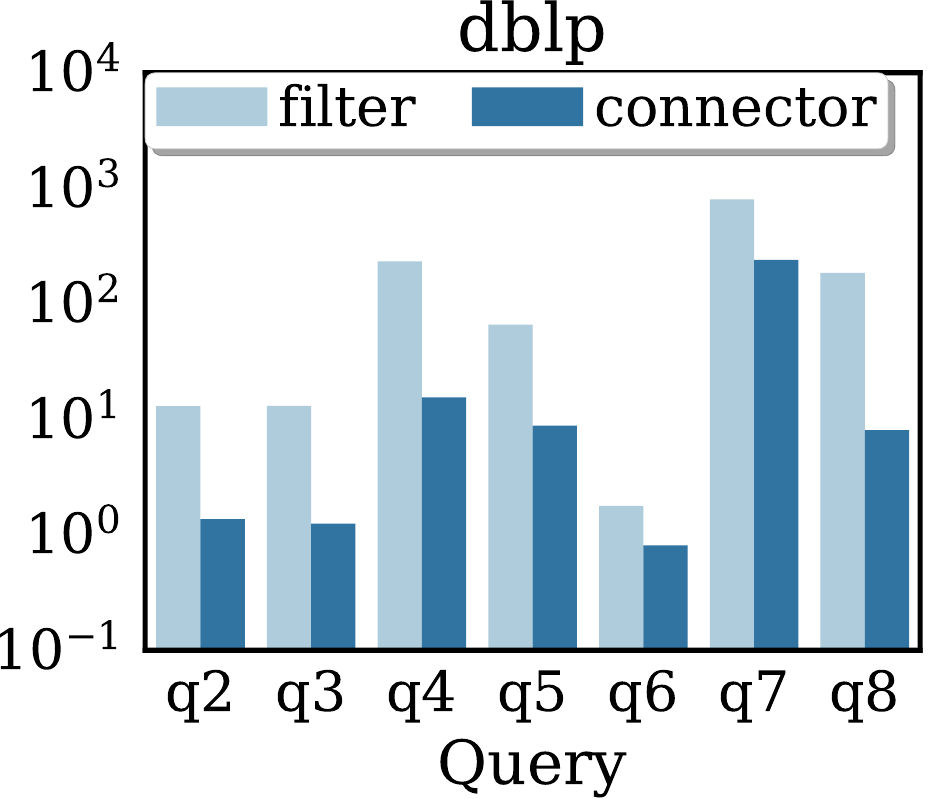}
  \includegraphics[width=0.216\textwidth]{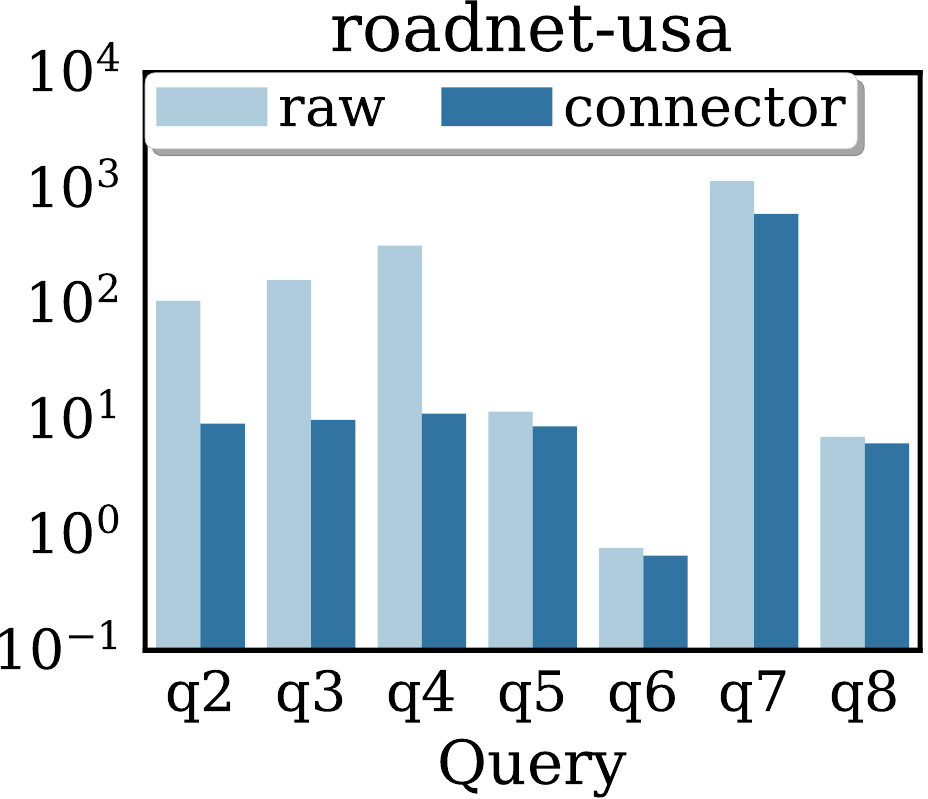}
  \includegraphics[width=0.216\textwidth]{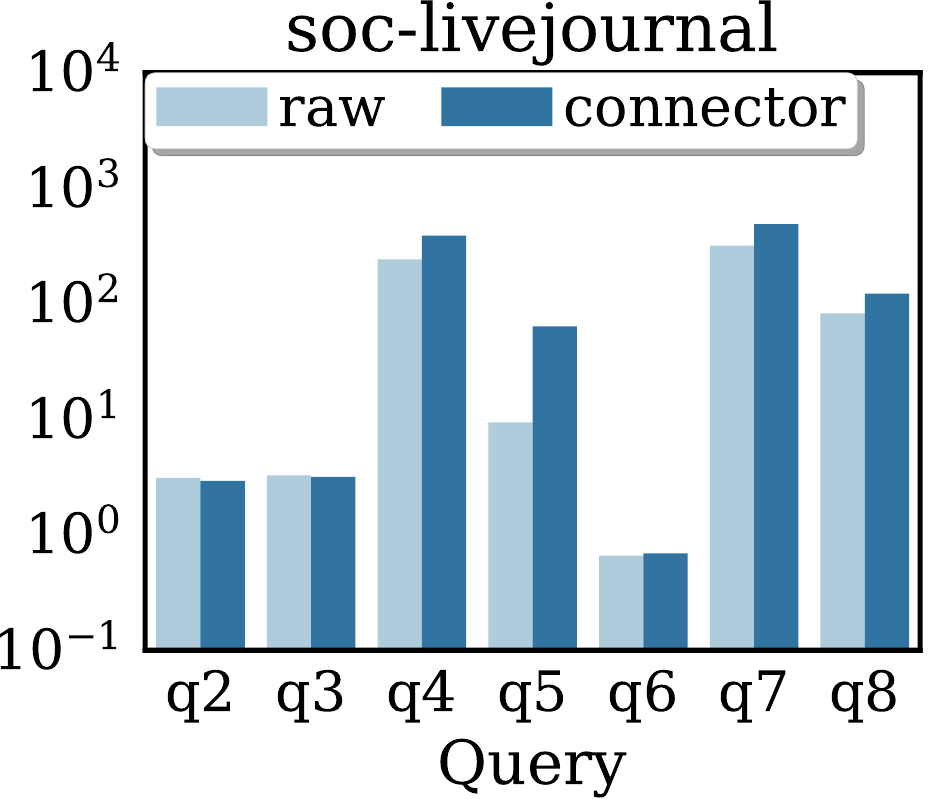}
    \caption{Total query execution runtimes over the graph after applying a
    \protect\vtype{sparsifier} view, and
    rewritten over a 2-hop \protect\vtype{spanner} view.
    \protect\vtype{Spanners} are \emph{job-to-job} (\texttt{prov}),
    \emph{author-to-author} (\texttt{dblp}), and \emph{vertex-to-vertex} for
    homogeneous networks \texttt{roadnet-usa} and \texttt{soc-livejournal}
    ($y$-axis is in log scale).}
 \label{fig:query_runtimes}
  \end{center}
\end{figure*}

\subsection{Query Runtimes}
\label{sec:exp_qr}
This experiment measures the difference in total query runtime for queries when executed from scratch over the filtered graph versus over an equivalent \vtype{spanner} view on the filtered graph. Figure~\ref{fig:query_runtimes} shows our results, with runtimes averaged over 10 runs, plotted in log scale on the $y$-axis.  Because the amount of data required to answer the rewritten query is up to orders of magnitude smaller (\autoref{sec:exp_sr}) in the heterogeneous networks, virtually every query over the \texttt{prov} and \texttt{dblp} graphs benefit from executing over the \vtype{spanner} view.  Specifically, Q2 and Q3 have the least performance gains (less than 2 times faster), while Q4 and Q8 obtain the largest runtime speedups (13 and 50 times faster, respectively).  This is expected: Q2 (``ancestors'') and Q3 (``descendants'') explore a $k$-hop ego-centric neighborhood in both the filtered graph and in the \vtype{spanner} view that is of the same order of magnitude.  Q4 (``path lengths'') and Q8 (``community detection''), on the other hand, are queries that directly benefit from the significant size reduction of the input graph. In particular, the maximum number of paths in a graph can be exponential on the number of vertices and edges, which affects both the count of path lengths that Q4 performs, as well as the label-propagation updates that Q8 requires.  Finally, we also observe that because \sysname creates \vtype{spanner} views through graph transformations that are engine-agnostic, these gains should be available in other systems as well.

For the homogeneous networks, we look at the total query runtimes over the raw graph and over a \emph{vertex-to-vertex} materialized $2$-hop \vtype{spanner}, which may be up to orders of magnitude larger than the raw graph (\autoref{sec:exp_vs}), as these networks have only one type of edge. Despite these differences in total sizes, a non-intuitive result is that the total query runtime is still linearly correlated with the order of magnitude increase in size for the network with a power-law degree distribution (\texttt{soc-livejournal}), while it decreases for path-related queries in the case of \texttt{roadnet-usa}. This is due to a combination of factors, including the larger fraction of long paths in \texttt{roadnet-usa}. Lastly, while the decision on which views to materialize heavily depends on the estimated view size (e.g., size budget constraints, and proxy for view creation cost), we note that these $2$-hop \vtype{spanner} views are unlikely to be materialized for the two homogeneous networks, due to the large view sizes predicted by our cost model, as shown in Figure~\ref{fig:view_size_estimates}. 

\section{Related Work}
\label{s:related}

\mypara{Views and language.}
Materialized views have been widely used in the relational setting to improve query runtime by amortizing and hiding computation costs~\cite{materializedviews}. This inspires our work, but the topological nature of graph views makes them new and different. The notion of graph views and algorithms for their maintenance was first introduced in\cite{graphviews} by Zhuge and Garcia-Molina in 1998, but since then there has been little attention in this area.  
With \sysname, we aim at providing a practical approach that can be
used to deal with large-scale graphs. In particular, \sysname addresses various problems related to graph views, including view enumeration, selection, as well as view-based query rewriting.
In this paper, we focus on extracting views for the graph traversal portion of our queries, since these are the most crucial for performance. An interesting avenue for future work is to address a combined analysis of the relational and graph query fragments, related to what \cite{sqlongraphs} proposes for a more restricted set of query types or \cite{chen2008graph} does for OLAP on graphs scenarios. The challenge is to identify and overcome the limits of relational query containment and view rewriting techniques for our hybrid query language.

Fan et al.~\cite{Fan2016} provide algorithms to generate views for speeding up fixed-sized subgraph queries, whereas \sysname targets traversal-type queries that can contain an arbitrary number of vertices/edges. They do not provide a system for selecting the best views to generate based on a budget constraint. Le et al.~\cite{LeDKLW11} present algorithms for rewriting queries on SPARQL views, but do not have a system for selecting views. Katsifodimos et al.~\cite{KatsifodimosMV12} present techniques for view selection to the improve performance of XML queries, which are limited to trees due to the structure of XML.


\mypara{Graph engines.}
\sysname is a graph query optimization framework that proposes a novel {\em constraint-based view inference} technique for materialized view selection and query rewriting. Although it is not a graph engine itself, in its current design it ultimately acts as one. We believe, however, that existing graph query execution engines may be able to leverage \sysname for query optimization. Therefore, we group existing graph data management approaches based on their main focus: \emph{specialized databases for graph queries}, including single-machine solutions, such as Neo4j~\cite{neo4j} and TinkerPop~\cite{tinkerpop}, and distributed approaches, such as Datastax Graph Engine~\cite{datastax}, TitanDB~\cite{titan}, CosmosDB~\cite{cosmosdb}; \emph{large scale graph analytics}, including Pregel~\cite{pregel},  GraphLab~\cite{graphlab}, GraphX~\cite{graphx}, Ligra~\cite{ShunB2013}, and EmptyHeaded~\cite{Aberger2017} (see~\cite{Yan2017,McCune2015} for surveys of graph processing systems); 
\emph{relational engines with graph support}, such as SQLServer 2017~\cite{sqlserver_graph}, Oracle~\cite{oracle_graph}, Spark GraphFrames~\cite{graphframes}, Agensgraph~\cite{agensgraph}; and \emph{graph algorithm libraries}, such as NetworkX~\cite{networkx}. 
Our approach is mostly independent of the underlying graph engine, and while we run our experiments on Neo4J, \sysname is directly applicable to any engine supporting Cypher~\cite{cypher} (or SQL+Cypher).

\mypara{RDF and XML.}
The Semantic Web literature has explored the storage and inference retrieval of RDF and OWL data extensively~\cite{antoniou2004semantic}. While most efforts focused on indexing and storage of RDF triples, there has also been work on maintenance algorithms for aggregate queries over RDF data~\cite{hung2005rdf}. While relevant, this approach ignores the view selection and rewriting problems we consider here, and it has limited applicability outside RDF. RDFViewS~\cite{rdfviews} addresses the problem of view selection in RDF databases. However, the considered query and view language support pattern matching (which are translatable to relational queries) and not arbitrary path traversals, which are crucial in the graph applications we consider. Similar caveats are present in prior work on XML rewriting~\cite{xmlviews2006, xmlviews2007}, including lack of a cost model for view selection.


\mypara{Graph summarization and compression.} Graph summarization is the task of finding a smaller graph that is representative of the original graph to speed up graph algorithms or queries, for graph visualization, or to remove noise~\cite{summarization}. Most related to our work is the literature on using summarization to speed up graph computations for certain queries, where the answers can either be lossless or lossy~\cite{Fan2012,Navlakha2008,Xirogiannopoulos2017,Tang2016,Chen2009}. As far as we know, prior work in this area has not explored the use of \vtype{spanners} and \vtype{sparsifiers} as part of a general system to speed up graph queries.

Rudolf et al.~\cite{Rudolf2013} describe several summarization templates in SAP HANA, which can be used to produce what we call graph views. However, their paper does not have a system that determines which views to materialize, and does not use the views to speed up graph queries.

There has been significant work on lossless graph compression to reduce space usage or improve performance of graph algorithms (see, e.g.,~\cite{BoldiV04,SDB2015,Maccioni2016}). This is complementary to our work on graph views, and compression could be applied to reduce the memory footprint of our views.

\section{Conclusions}
\label{s:conclusions}

We presented \sysname, a graph query optimization framework that employs materialization to efficiently evaluate queries over graphs. Motivated by the fact that many application repeatedly run similar queries over the same graph, and that many production graphs have structural properties that restrict the types of vertices and edges that appear in graphs, \sysname automatically derives graph views, using a new {\em constraint-based} view enumeration technique and a novel cost model that accurately estimates the size of graph views. We show in our experiments that queries rewritten over some of these views can provide up to \speedup times faster query response times. Finally, the query rewriting techniques we have proposed are engine-agnostic (i.e., they only rely on fundamental graph transformations that typically yield smaller graphs, such as path contractions) and thus are applicable to other graph processing systems.


\bibliographystyle{IEEEtran}
\bibliography{IEEEabrv,paper}



\appendix

\section{Example Summarizer View Templates}
\label{ss:additional-view-templates}

In~\autoref{s:venum}, we presented view templates in Prolog for \vtype{spanner} views. Similarly, in~\autoref{listing:extraction_rules_b}, we provide example view templates for \protect\vtype{sparsifier} views.

\begin{listing}[htb!]
\prologcode{listings/extraction_rules_b.pl}
  \caption{Example view template definitions for \protect\vtype{sparsifiers}.}
\label{listing:extraction_rules_b}
\end{listing}

\section{Constraint Mining Rules}
\label{ss:additional-constraint-mining-rules}

As discussed in \autoref{s:venum}, \sysname uses a set of constraint mining rules to extract {\em implicit} constraints from both schema and query graphs. In~\autoref{listing:extraction_rules_a}, we provide the Prolog definitions for some {\em query} constraint mining rules (for an example schema constraint mining rule, see \autoref{s:venum}) used in \sysname, such as path constraints over query graphs. In addition, \autoref{algo:constraint_mining_example} provides a procedural version of the \texttt{schemaKHopPath} example schema constraint mining rule of~\autoref{listing:structural_properties}. \KK{Make sure to include a quick description of the algorithm (space permitted).}

\vspace{3mm}

\begin{listing}[htb!]
\prologcode{listings/extraction_rules_a.pl}
\caption{Example query constraint mining rules.}
\label{listing:extraction_rules_a}
\end{listing}

\vspace{3mm}

\begin{algorithm}[htb!]
\caption{Procedural version of \texttt{schemaKHopPaths} constraint mining Prolog rule}
\label{algo:constraint_mining_example}
\begin{algorithmic}[1]
\CommentLine{Procedural version of one the declarative constraint mining}
\CommentLine{programs that bound the search space for valid candidate}
\CommentLine{views~(\autoref{listing:structural_properties} and \autoref{ss:strschema})}
\Function{k\_hop\_schema\_paths}{schema\_edges, paths, k, curr\_k}
  \If {curr\_k == 0}
    \Return [p {\bf for} p $\in$ paths {\bf if} len(p) == k]
  \EndIf
  \If {k == curr\_k}
    \State{new\_paths $\gets$ [[e] {\bf for} e $\in$ schema\_edges]}
    \State \Return k\_hop\_schema\_paths(schema\_edges, new\_paths, k, k-1)
  \EndIf
  \State {new\_paths $\gets$ []}
  \For {\{i, path\} $\in$ paths}
      \State {src, dst $\gets$ path[0][0], path[-1][1]}
      \For {\{j, edge\} $\in$ schema\_edges}
          \CommentLine{Add edge to the end of the path.}
          \If {dst == edge[0]} {new\_paths.append(path + [edge])}\EndIf
          \CommentLine{Add edge to the front of the path.}
          \If {src == edge[1]} {new\_paths.append([edge] + path)}\EndIf
      \EndFor
  \EndFor
  \CommentLine{Step omitted: duplicate paths removal.}
  \CommentLine{Fix-point: only include paths that grew this round.}
  \State {paths $\gets$ [p {\bf for} p $\in$ new\_paths {\bf if} len(p) == (k-curr\_k+1)]}
    \State \Return {\textproc{k\_hop\_schema\_paths}(schema\_edges, paths, k, curr\_k-1)}
\EndFunction
\end{algorithmic}
\end{algorithm}

\vspace{3mm}

\section{Degree Distributions}
\label{ss:degree-distributions}

\begin{figure}[htb!]
\minipage{0.22\textwidth}
  \centering
  \includegraphics[width=\columnwidth]{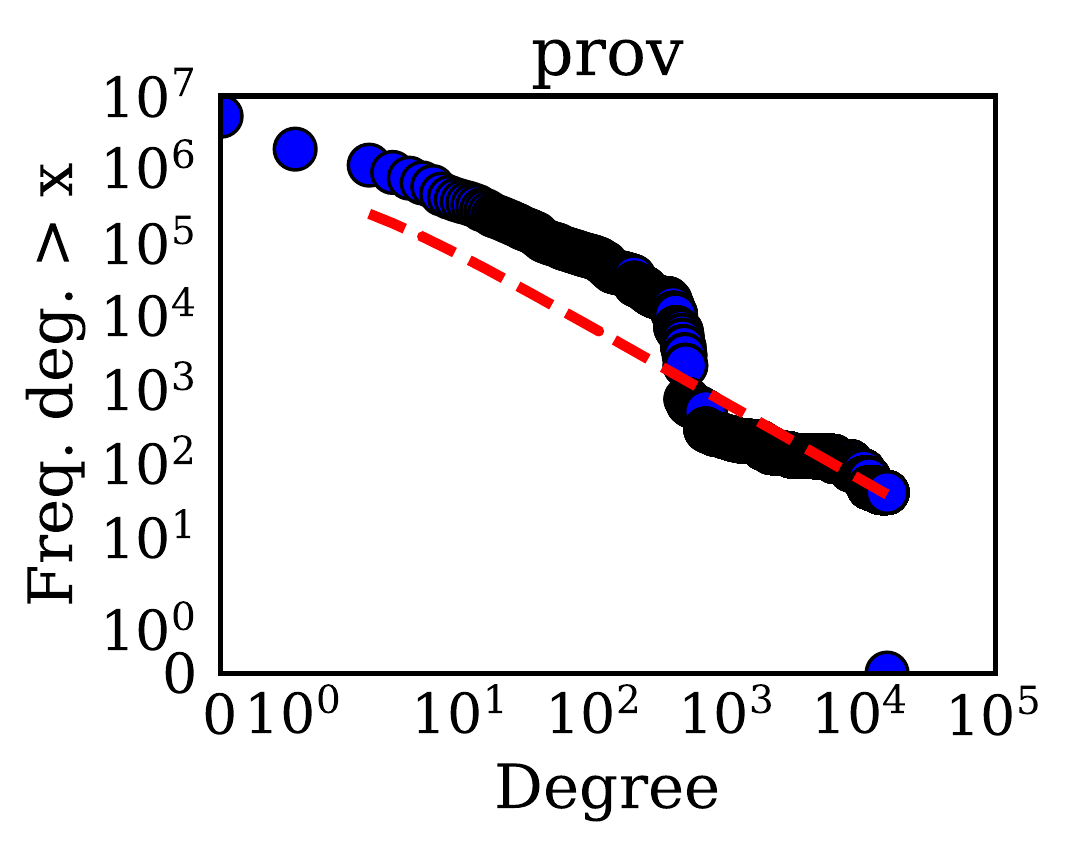}
\endminipage
\minipage{0.22\textwidth}
  \centering
  \includegraphics[width=\columnwidth]{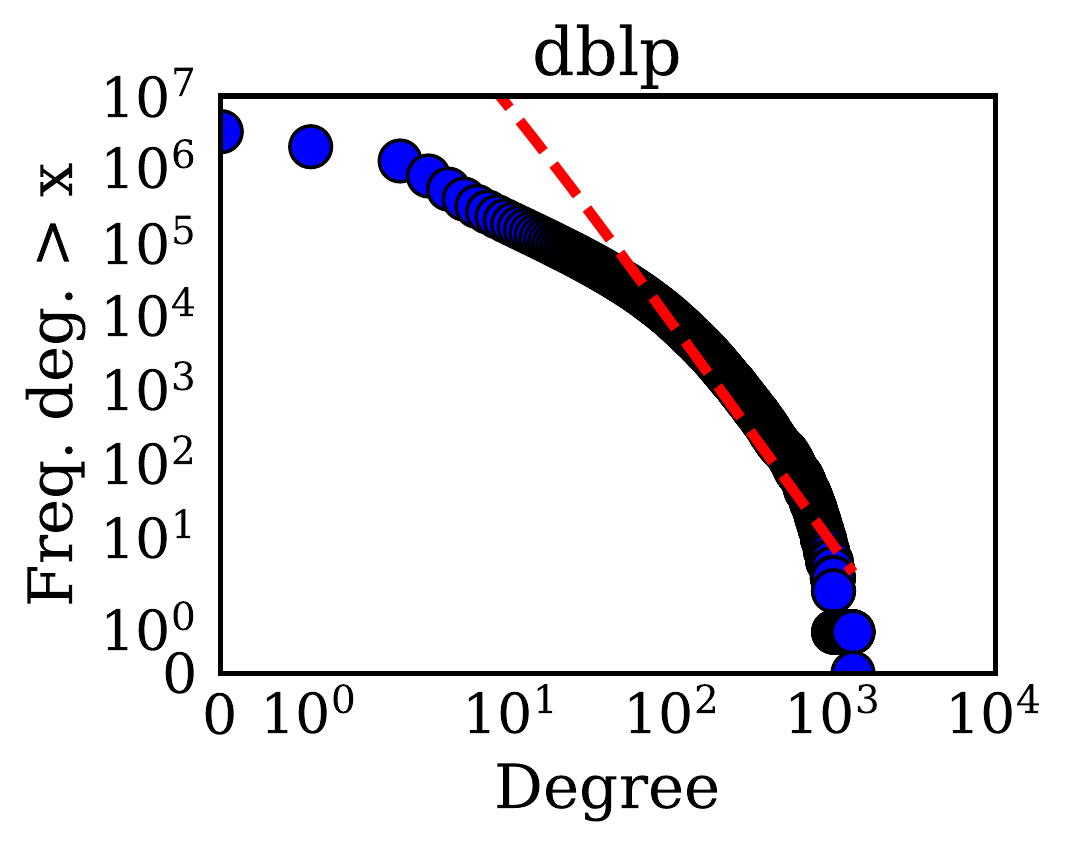}
\endminipage\hfill
\minipage{0.22\textwidth}
  \centering
  \includegraphics[width=\columnwidth]{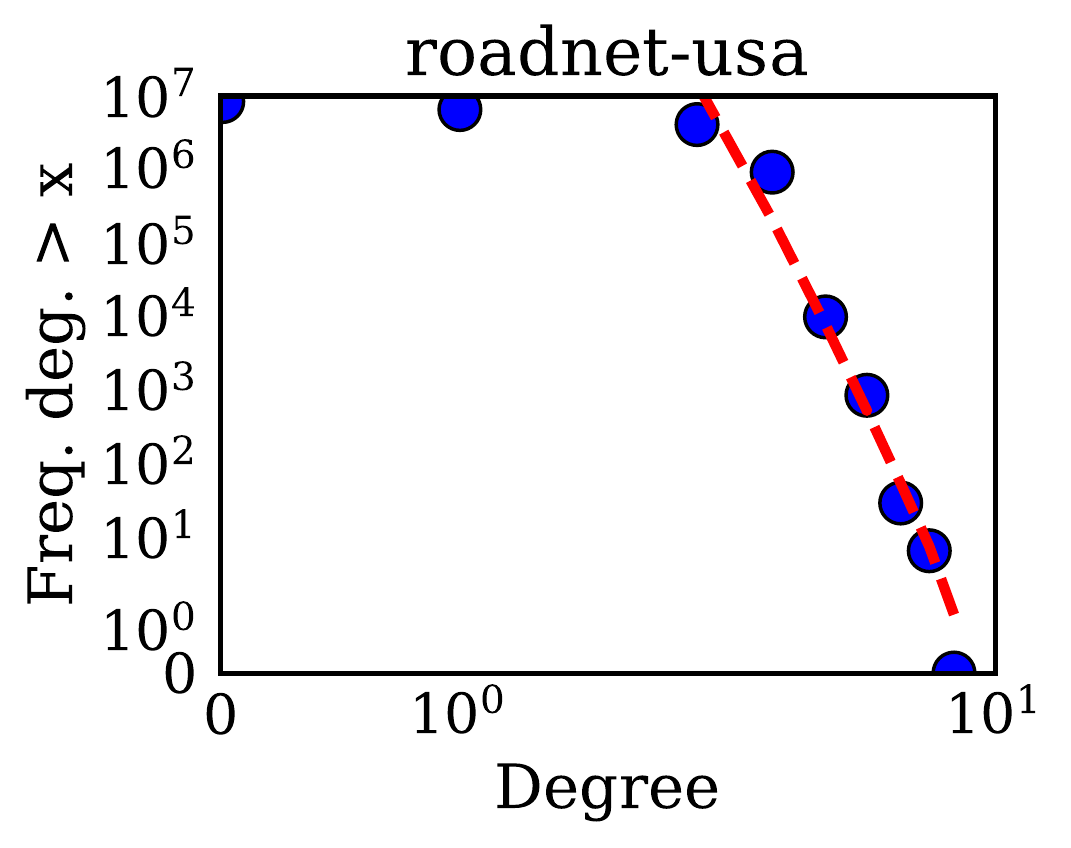}
\endminipage
\minipage{0.22\textwidth}
  \centering
  \includegraphics[width=\columnwidth]{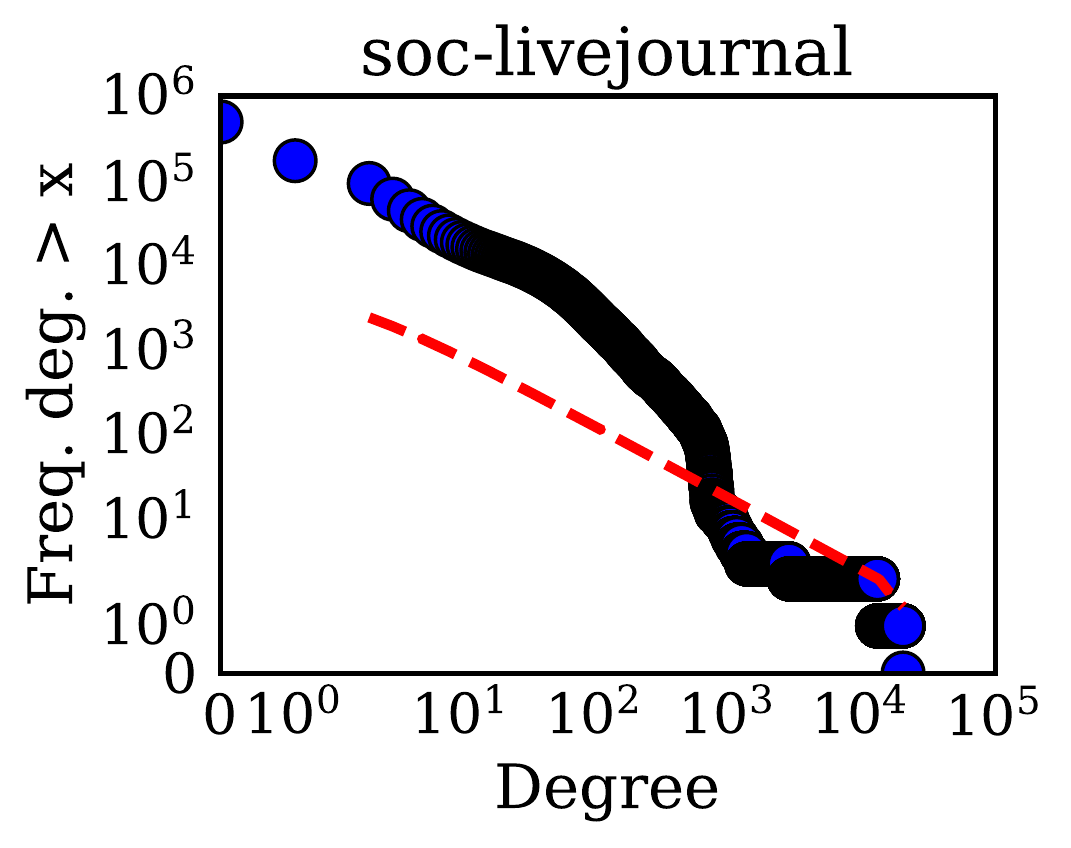}
\endminipage
  \caption{Degree distribution log-log CCDF plots, together with the best-fit power-law exponent (linear on log-log scale) for all vertices in each dataset.}
\label{fig:deg_dist}
\end{figure}

\autoref{fig:deg_dist} shows the degree distributions of the different 
 graphs used in our evaluation.  As expected, vertex degrees in all but the
 road network dataset are roughly modeled by a power-law distribution,
 as evidenced by a goodness-of-linear-fit on log-log plot of the complementary cumulative distribution function (CCDF).

\end{document}